\documentclass[twocolumn,superscriptaddress,amsmath,amssymb,aps,prl,floatfix]{revtex4-2}

\usepackage{graphicx}
\usepackage{bm}

\usepackage[utf8]{inputenc}
\usepackage[T1]{fontenc}

\usepackage{braket}
\usepackage{xcolor}

\usepackage[normalem]{ulem}

\begin{document}


\title{Parity-doublet spin interactions in ultracold polyatomic molecules}

\author{Junheng Tao}
\author{Paige Robichaud}
\author{Christian Hallas}
\author{Giseok Lee}
\author{John M. Doyle}
\email{jdoyle@g.harvard.edu}
\affiliation{Department of Physics, Harvard University, Cambridge, MA 02138, USA}
\affiliation{Harvard-MIT Center for Ultracold Atoms, Cambridge, MA 02138, USA}

\date{\today} 

\begin{abstract}
Precision measurements exploit internal molecular structure to search for physics beyond the Standard Model~\cite{acme2018improved, kozyryev2017precision, anderegg2023quantum, norrgard2019nuclear, demille2024quantum, roussy2023improved}.
Reaching the entanglement-enhanced regime requires preparing and preserving correlated molecular states within the same manifold used for sensing~\cite{zhang2023quantum, bilitewski2021dynamical, pezze2018quantum, holland2026creating}.
Parity-doublet states are central to proposals for precision measurements with polyatomic molecules because they enable large molecular polarization for enhanced sensitivity, and provide internal state co-magnetometry to suppress systematic errors~\cite{acme2018improved, kozyryev2017precision, anderegg2023quantum, demille2024quantum, roussy2023improved, takahashi2026engineered}.
Here we report the observation and control of resonant dipolar exchange interactions in a parity-doublet sensing manifold of ultracold polyatomic molecules.
Using laser-cooled CaOH molecules in a crossed optical dipole trap, we encode a pseudo-spin in the $\ell$-type parity-doublet states and observe two complementary signatures of interactions:
a density-dependent Ramsey contrast decay and a spin-projection-dependent mean-field precession whose sign reverses with the initialized population imbalance.
Both observables are well described by a common discrete truncated Wigner model~\cite{schachenmayer2015dynamics, schachenmayer2015many, zhu2019generalized} using independently calibrated experimental parameters.
We further Floquet-engineer the interaction anisotropy and observe suppressed many-body dephasing near the SU(2)-symmetric Heisenberg point, identifying a protected regime relevant for spin squeezing and precision measurement with dense molecular ensembles~\cite{norcia2018cavity, davis2020protecting, koyluoglu2025squeezing}.
Our results establish parity-doublets as a coherent, interacting, and programmable molecular degree of freedom, connecting the polyatomic precision-measurement platform to dipolar many-body physics and opening a route to entanglement-enhanced tests of fundamental physics.
\end{abstract}

\maketitle

Molecules serve as quantum sensors of fundamental physics because their internal structure can amplify tiny symmetry-violating effects.
A particularly powerful example is the use of closely spaced opposite-parity states, which weak electric fields can mix to polarize the molecule, while reversals within the doublet enhance suppression of systematic errors by distinguishing molecular-frame signals from laboratory-frame systematic shifts.
Parity-doublet states are already a key feature of the best probes of CP-violating (CP-v) physics beyond the Standard Model (BSM), and are very promising for future precision measurements and quantum simulations~\cite{acme2018improved, kozyryev2017precision, anderegg2023quantum, norrgard2019nuclear, demille2024quantum, roussy2023improved}.
Linear polyatomic molecules host these parity-doublets in their bending vibrational modes~\cite{hallas2023optical}.
In polyatomic molecules, parity-doublet structures have been proposed for CP-v BSM physics searches through the electron electric dipole moment (eEDM) and CP-v nuclear magnetic quadrupole moments, both at scales $>10~\mathrm{TeV}$, and nuclear-spin-dependent parity violation~\cite{anderegg2023quantum, zhang2023quantum, kozyryev2017precision, norrgard2019nuclear, demille2024quantum}.
Recent experiments have prepared polyatomic molecules in eEDM-sensitive states, establishing them as promising BSM probes~\cite{anderegg2023quantum, takahashi2026engineered}.
However, these measurements have so far relied on independent molecules.
For parity-doublets to become useful many-body resources, entanglement-enhanced precision measurements call for coherent interactions between molecules that act within the same metrologically important manifold~\cite{zhang2023quantum}.

Ultracold molecules offer the ingredients needed to enter this many-body regime.
Rotational states in ultracold diatomic molecules have been used to encode qubits, with their electric dipole-dipole interactions providing a route to analog quantum simulations of spin models and quantum computations~\cite{yan2013observation, bao2023dipolar, holland2023demand, gregory2024second, li2023tunable, christakis2023probing, micheli2006toolbox, gorshkov2011quantum, picard2025entanglement, ruttley2025long, demille2002quantum, carr2009cold,bohn2017cold, cornish2024quantum}.
Polyatomic molecules provide an enabling architecture to connect the many-body toolbox to precision measurement: their complex vibrational and rotational structure naturally support a plethora of parity-doublets while remaining compatible with laser cooling and optical trapping~\cite{kozyryev2017sisyphus, vilas2022magneto, hallas2023optical}.
These parity-doublets can carry the symmetry-violating signal and are themselves electrically coupled, so interactions act directly within the sensing manifold.
At zero electric field, the two states in a parity-doublet are near degenerate and share identical quantum numbers apart from their inversion parity.
This near-identity suppresses differential sensitivity to magnetic and optical perturbations, enabling second-scale qubit coherence in optically trapped polyatomic molecules~\cite{robichaud2026parity}. 
These ingredients point toward a new regime in which the metrological advantages of molecular parity-doublet states can be extended by the programmable interactions of ultracold dipolar many-body systems.

In this work, we establish parity-doublet states as a programmable interacting degree of freedom in ultracold polyatomic molecules.
We realize and control coherent resonant dipolar exchange interactions between CaOH molecules directly within the $\ell$-type parity-doublet manifold of the $\tilde{X}^2\Sigma^+(010)$ vibrational bending mode relevant to precision measurement.
Molecular thermal motion generates spatially disordered, long-range exchange couplings, realizing an itinerant XY spin model.
As a first manifestation of interactions, a rapid collective spin decay is observed as a function of molecular density.
We characterize the system spin Hamiltonian by initializing different spin projections and measuring the spin precession, revealing a mean-field one-axis-twisting (OAT) interaction that scales linearly with density and reverses direction with the population imbalance.
We then use Floquet control to tune the interaction anisotropy and observe strongly suppressed dephasing near the SU(2)-symmetric Heisenberg point, despite the inhomogeneous couplings. 
This identifies a protected interacting regime for precision measurements at high molecule number, mitigating the interaction-induced coherence penalty that otherwise grows with molecular density.
The ability to switch between anisotropic exchange that generates correlations and isotropic interaction that preserves collective coherence establishes parity doublets as a programmable many-body resource, and offers a route to generate-and-protect protocols for molecular quantum sensing.

We create ultracold CaOH molecules by laser slowing and cooling a molecular beam from a cryogenic buffer gas source and loading them into a red-detuned radio frequency (rf) magneto-optic trap (MOT)~\cite{vilas2022magneto, lasner2025magneto}. 
They are then transferred to a compressed dc blue-detuned conveyor-belt MOT to enhance the density~\cite{hallas2026high, li2025conveyor}, further cooled~\cite{hallas2023optical}, and loaded into a small volume of a crossed optical dipole trap (cODT). 
Approximately $90$ molecules at $T= 60~\mathrm{\mu K}$ are confined by two $\lambda=1064$~nm lasers that intersect at $45^\circ$ to create the cODT.
One of the cODT laser beams has an elliptical aspect ratio of 5, resulting in an oblate crossed trap with motional frequencies $\omega/2\pi= (1.3(1), 3.5(1), 11.5(2))~\mathrm{kHz}$ at trap depth $U_0/k_B= 1.3(2)~\mathrm{mK}$.

Molecules in the cODT are optically pumped into the excited vibrational bending mode $\tilde{X}^2\Sigma^+(010)$ and prepared into a single parity-doublet stretched hyperfine state $\ket{N=1, J=3/2, F=2, m_F = 2, p=+}$ using combined optical and microwave pumping as well as an rf adiabatic sweep~\cite{robichaud2026parity}.
The polarization of the two dipole beams is set at the magic condition to optimize the single-molecule parity-doublet coherence, which is limited by differential AC Stark shifts from the trap light.
After the state preparation, the bias magnetic field is ramped up to $B_z= 30~\mathrm{G}$ to further enhance the coherence time and define the quantization axis $z$.

\begin{figure}[!]
\centering
\includegraphics{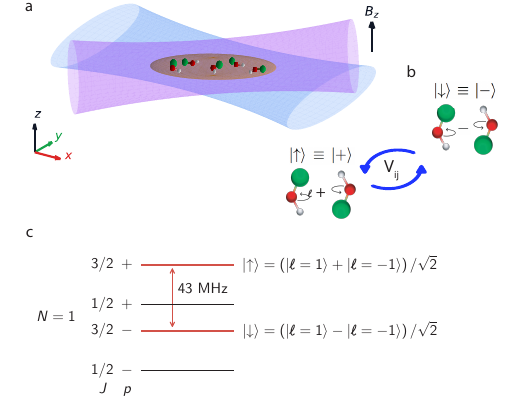}
\caption{Experimental setup.
{\bf a} Itinerant polyatomic molecules in a crossed optical dipole trap, which provides tighter confinement in $z$.
The anisotropy enhances the OAT mean interaction strength.
{\bf b} The resonant dipolar exchange interaction happens between a pair of molecules that respectively reside in the two parity eigenstates made by superpositions of opposite bending angular momentum states labeled by projections $\ell=\pm 1$.
The lab quantization axis is defined by a strong bias magnetic field along $z$.
{\bf c} The $N=1$ rotational state diagram of $\tilde{X}^2\Sigma^+(010)$ in CaOH, which contains our parity-doublet manifold. 
The two states that comprise our qubit system are ones of opposite parity, $|+\rangle ~{\rm{and}}~ |-\rangle $, which we label
$\ket{\uparrow}\equiv\ket{+}$ and 
$\ket{\downarrow} \equiv\ket{-}$.
}
\label{fig:setup}
\end{figure}

Our two-level system is formed by the stretched hyperfine states of the opposite parity components of the $\ell$-type doublet.
We define $\ket{\uparrow}\equiv\ket{N=1,J=3/2,F=2,m_F=2,p=+}$ and $\ket{\downarrow}\equiv\ket{N=1,J=3/2,F=2,m_F=2,p=-}$, which are separated by a small parity-doublet splitting and share the same angular-momentum quantum numbers except parity, as shown in Fig.~\ref{fig:setup}\textbf{c}.
Here, $N$ is the rotational quantum number, $J$ includes the electron spin, $F$ is the total angular momentum including nuclear spin, $m_F$ is the projection of $F$ along the quantization axis $z$, and $p$ denotes the parity.
Because the two states have nearly identical angular-momentum structure, differential magnetic moments and differential AC Stark shifts are very small; residual trap sensitivity is a parity-dependent AC Stark shift arising from Renner–Teller mixing in the $\tilde{A}^2 \Sigma(010)$ excited state~\cite{robichaud2026parity}.
At the same time, these states of opposite parity have a large transition electric dipole moment $d_{\uparrow\downarrow}=\bra \uparrow d \ket{\downarrow}=0.73~\mathrm{D}$.
Coherent rf pulses resonant with the parity-doublet transition prepare arbitrary pseudo-spin states on the Bloch sphere, with transverse spin components $s_{\perp}$ corresponding to coherence between $\ket{\uparrow}$ and $\ket{\downarrow}$ and $s_z$ corresponding to their population imbalance.

At zero electric field, parity remains a good quantum number, and the molecule has no lab-frame dipole moment.
In this case, the electric transition dipole moment between the two opposite-parity states gives rise to resonant dipolar exchange interactions between molecules, described by an XY spin Hamiltonian,
\begin{equation}
\label{eq: H_XY}
\hat{H} =\frac{1}{2}\sum_{i\neq j} V(\mathbf{r}_i-\mathbf{r}_j)
\left(s_i^x s_j^x+s_i^y s_j^y\right)
\end{equation}
where
$    V(\mathbf{r})= 2d_{\uparrow\downarrow}^{\,2} \left(1-3\cos^2\alpha\right) / (4\pi\epsilon_0 r^3)$
is the anisotropic dipole-dipole interaction strength between $i, j$ and is determined by the instantaneous molecular separation $r$ and the angle $\alpha$ between the intermolecular axis and the quantization field.
In the cODT, the molecules undergo thermal itinerant motion, sampling a time-dependent and spatially disordered set of exchange couplings. This produces two experimentally measurable signatures: an inhomogeneous decay of the collective transverse spin length and a mean-field precession of the transverse spin whose rates depend on the initialized spin projection.

We model these many-body dynamics through the discrete truncated Wigner approximation (DTWA)~\cite{schachenmayer2015many, schachenmayer2015dynamics, zhu2019generalized, bilitewski2021dynamical}. 
The initial molecular quantum spin state is sampled over an ensemble of classical spin configurations, which is then evolved under the time-varying dipolar exchange couplings generated by thermal molecular trajectories in the trap potential.
The simulations use independently calibrated trap frequencies, temperature, and molecule number.
Observables are obtained by averaging the collective spin over many sampled spin configurations and molecular trajectories.

Experimentally, we calibrate the parity-doublet spin coherence in the absence of interactions with two approaches: either (i) turning off one of the two cODT beams, or (ii) reducing the loading of molecules to $\sim 10\%$ of the maximum number.
In the single ODT beam approach (i), we carefully cancel the lab electric fields and optimize the trap polarization at the magic angle condition.
We achieve a bare coherence time of $T_2^*=14(2)~\mathrm{ms}$ at a trap depth of $0.9(2)~\mathrm{mK}$, and by performing an XY16 dynamical decoupling multi-pulse sequence~\cite{gullion1990new}, we extend $T_2^*$ by more than a factor of 60, beyond the vibrational bending mode $1/e$ radiative lifetime $T_r \sim350~\mathrm{ms}$~\cite{vilas2023blackbody}. 
The time interval $\tau_p=25~\mathrm{\mu s}$ between each $\hat{X}$ or $\hat{Y}$ pulse is chosen to match the cutoff set by the largest trap frequency $\omega_z$ and its harmonics.
In case (ii), the low molecule density condition allows us to measure the bare coherence time, $T_2^* = 2.5(4)~\mathrm{ms}$, and the XY16 extended coherence time, $T_{2-16} = 250(30)~\mathrm{ms}$, well beyond the time scale observed from dipolar interaction induced phase changes.

By decomposing the pairwise interaction $V_{ij} = \bar V + \delta V_{ij}(t)$, the Hamiltonian~\eqref{eq: H_XY} has a mean-field contribution, through the definition of collective spin $S_\alpha=\sum_i s_i^\alpha$ and the ensemble-averaged interaction strength $\chi=\bar V/2$
\begin{equation}
\label{eq: H_coll}
\hat{H}_{coll} = \chi S^2-\chi S_z^2+\mathrm{const.}
\end{equation}
The first collective $S^2$ term tends to preserve the total spin length $|S|$; the nonlinear dynamics are generated by the one-axis-twisting term $-\chi S_z^2$, which shears the collective-spin quasiprobability distribution.
This produces a coherent loss of contrast and, at later times, is expected to generate spin squeezing and metrologically useful entanglement~\cite{kitagawa1993squeezed, wineland1994squeezed, pezze2018quantum}.
In addition, as thermal molecules ergodically traverse the trap, the pairwise couplings $\delta V_{ij}(t)$ fluctuate in time and vary between molecular trajectories.
Equivalently, each molecule samples a different accumulated interaction phase set by its local environment. The resulting distribution of interaction-induced phases broadens with molecular density, leading to the observed density-dependent dephasing.

We measure such dephasing by performing a Ramsey pulse sequence.
The first $\pi/2$ rf pulse prepares a transverse polarized spin state $(\ket{\uparrow}+\ket{\downarrow})/\sqrt{2}$ for each molecule.
The following blocks of XY16 dynamical decoupling pulses during the interaction time $T$ act as a single-body echo, and a last $\pi/2$ readout pulse phase is scanned to extract the Ramsey fringe contrast.
We use a parity-selective optical detection scheme to measure the population in $\ket{\uparrow}$ and repeat the measurement with different densities of molecules.
We observe that contrast decay fits well to an exponential function of $T$, where the characteristic decay rate $\Gamma$ becomes faster at higher densities.

\begin{figure}[!]
\centering
\includegraphics{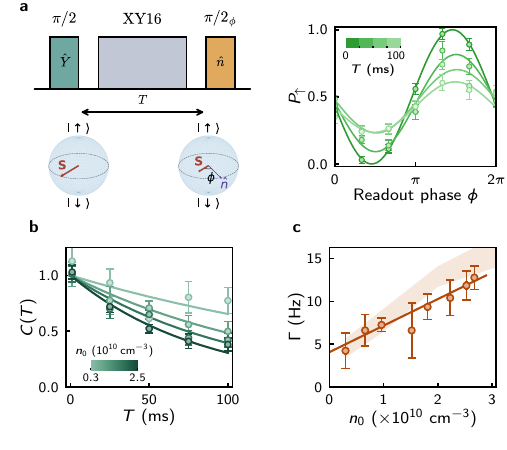}
\caption{Parity-doublet spin dephasing.
{\bf a} The Ramsey pulse sequence comprises a $\pi/2$ pulse that prepares the spin on the equator along $\hat{X}$, and a final $\pi/2$ that reads out the projection of $S_{\perp}$ along a certain axis $\hat{n}$ defined by the pulse phase $\phi$.
During the interval $T$ between these pulses, the spin system evolves under Hamiltonian $\eqref{eq: H_XY}$, while dynamical decoupling pulses suppress the single-particle decoherence arising from the trap shift.
On the right is an example of the measured fringes at different $T$ indicates fast contrast decay at $1, 25, 50, 100~\mathrm{ms}$.
Error bars represent $68\%$ confidence intervals from bootstrapping.
The contrast $C = S_\perp (T) / S_\perp (0)$ is extracted by fitting each Ramsey fringe to a sinusoidal function. 
{\bf b} The Ramsey contrast decay at different initial molecular peak densities $n_0= 0.3, 0.96, 1.81, 2.53 \times 10^{10} \mathrm{cm^{-3}}$.
For each density, we fit an exponential decay $C(T) = C_0 \exp(-\Gamma t)$.
{\bf c} Decay rate $\Gamma$ as a function of density.
The solid line is a linear fit with an intercept that indicates the remaining single-particle decoherence rate baseline.
Error bars in \textbf{b, c} are 1~standard error from the fits.
The shaded region represents the prediction from the DTWA simulations with $25\%$ uncertainty in $n_0$.
}
\label{fig:dephasing}
\end{figure}

The contrast decay is sensitive to the spread of the interaction-induced phase across the ensemble. 
In an additional probe of the Hamiltonian~\eqref{eq: H_coll}, we deliberately prepare a population imbalance between the two parity-doublet states.
In this case, linearizing the OAT term $-\chi S_z^2$ gives an effective mean-field shift $-2\chi \langle S_z\rangle S_z$ that causes the collective spin to precess about the $z$ axis at a rate proportional to both $\chi$ and the initialized spin projection $\langle S_z \rangle$.
The precession rate is expected to scale linearly with molecular density and to reverse sign when the population imbalance is reversed.
Measuring this dependence distinguishes coherent interaction-driven evolution from generic density-dependent dephasing.

\begin{figure}[tb!]
\centering
\includegraphics{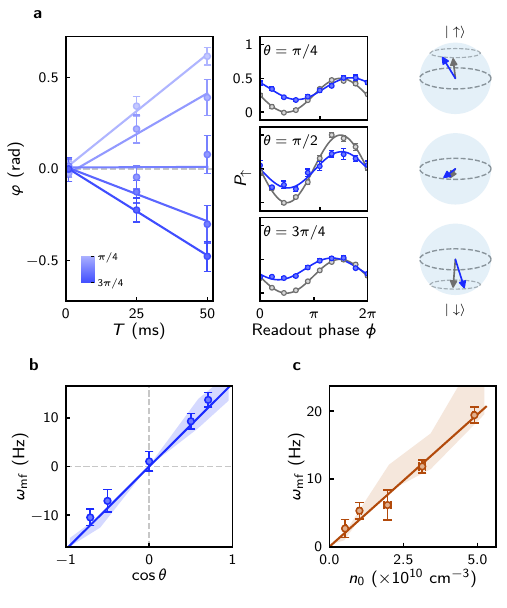}
\caption{Parity-doublet spin precession.
{\bf a} Collective spin precession phase $\varphi$ as function of $T$ for different initial angle $\theta$.
Colors from light to dark correspond to increasing $\theta= \pi/4, \pi/3, \pi/2, 2\pi/3, 3\pi/4$.
The initial spin is prepared by a pulse with area $\theta$ from $\ket{\uparrow}$, which, after time $T$, is rotated by the OAT interaction by azimuthal angle $\varphi(T)$.
Dynamical decoupling is implemented during $T$.
We again extract $\varphi$ by varying the last readout $\pi-\theta$ pulse phase $\phi$ and fitting to a sinusoidal.
The middle panel presents the normalized Ramsey fringes at $T=0$ (grey) and $50~\mathrm{ms}$ (blue).
Error bars represent $68\%$ confidence intervals from bootstrapping.
{\bf b} Mean-field precession rate $\omega_{mf}$, extracted from linear fits to the early-time phase evolution in \textbf{a}, as a function of the initialized spin projection $\cos\theta$.
In \textbf{a, b}, $n_0=3.1(1)\times 10^{10}~\mathrm{cm^{-3}}$.
{\bf c} $\omega_{mf}$ is linearly dependent on $n_0$ at $\theta=\pi/4$.
In \textbf{b, c} the shaded region represents DTWA simulation predictions.
}
\label{fig:precession}
\end{figure}

We prepare such an imbalanced state $\cos (\theta/2) \ket{\uparrow} + \sin (\theta/2) \ket{\downarrow}$ in each molecule through a variable initializing rf pulse with pulse area $\theta$. 
While dynamical decoupling pulses during $T$ keep the Ramsey fringe phase intact, the mean-field shift drives a precession rate $\omega_{mf} = \chi N \cos \theta$.
As shown in Fig.~\ref{fig:precession}, the accumulated phase is approximately linear in time over the measured evolution window, allowing us to extract $\omega_{\mathrm{mf}}$. 
The sign and magnitude of $\omega_{\mathrm{mf}}$ depend on the initialized spin projection, reversing when $\langle S_z\rangle$ is reversed and vanishing near the equator. 
Moreover, for a fixed spin projection, the precession rate increases linearly with molecular density.

Taken together, the dephasing and precession measurements probe complementary components of the same exchange interaction. 
The Ramsey contrast decay reveals the distribution of local interaction-induced phases, while the spin-projection-dependent precession provides a signed measurement of the ensemble-averaged interaction field. 
Their simultaneous agreement with the same DTWA calculation identifies the observed dynamics as resonant dipolar exchange within the parity-doublet manifold.

\begin{figure}[!]
\centering
\includegraphics{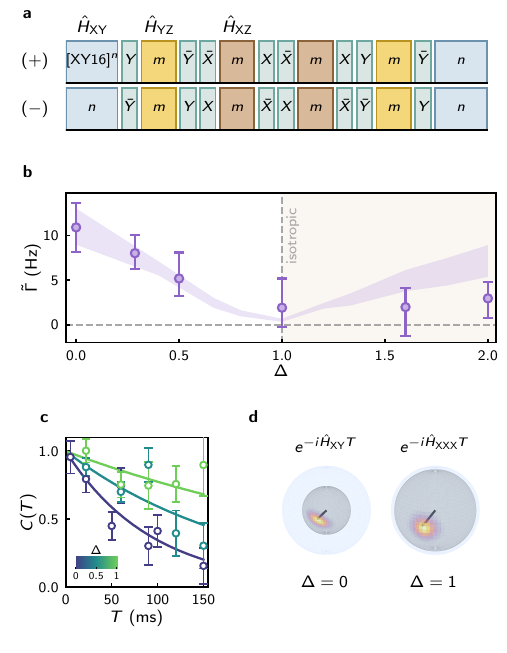}
\caption{Symmetry-protected collective coherence in parity-doublets.
{\bf a} A unit block of the Floquet pulse sequence contains frame-switching pulses (green) and interaction intervals (blue, yellow, and brown blocks).
Here $Y$ and $\bar{Y}$ denote $\pi/2$ rotations about $y$ and $-y$ respectively, with BB1 composite pulses.
During the interaction intervals, the spin system effectively evolves under Hamiltonians $\hat{H}_{AB}\propto AA+BB$ depending on the frame selection.
XY16 pulses are embedded within each interaction interval.
The $+$ and $-$ cycles use opposite frame rotations and are alternated to cancel leading-order pulse errors from the frame-switching pulses.
{\bf b} Engineered many-body dephasing rate $\tilde{\Gamma}$ as a function of the XXZ anisotropy $\Delta$.
The purple shaded region represents the prediction from the DTWA simulations.
Error bars represent $68\%$ confidence intervals from bootstrapping.
{\bf c} Representative Ramsey contrast decays $C(T)$ for selected values of $\Delta=0, 0.5, 1$ (dark to light).
Lines are exponential fits used to extract $\tilde{\Gamma}$.
{\bf d} DTWA snapshots of the collective spin distribution under the native XY interaction and the engineered isotropic XXX interaction at $T=35~\mathrm{ms}$.
Spatially inhomogeneous XY couplings broaden the distribution and reduce $S_\perp$, whereas the SU(2)-symmetric XXX interaction preserves the fully polarized collective-spin state despite the coupling disorder.
}
\label{fig:floquet}
\end{figure}

Having established coherent exchange interactions in the parity-doublet manifold, we now ask whether they can be engineered to overcome the observed density-dependent dephasing and yield an advantage for many-body metrology.
The native long-range XY interaction plays a dual role:
it supplies the nonlinear spin dynamics needed to generate correlations, while its spatially disordered couplings degrade the collective coherence required for precision measurement, even though itinerant motion partially averages this disorder.
For metrology, this usually presents a tradeoff: increasing the molecule number improves projection-noise-limited sensitivity~\cite{nasir2026order}, yet at fixed volume it also raises the dephasing rate~\cite{milner2025coherent}.
Symmetry offers a route to suppress this dephasing.
In the isotropic Heisenberg limit,
$\hat{H}_{\mathrm{XXX}}= \sum_{i\neq j} V_{ij}~
\mathbf{s_i} \cdot \mathbf{s_j}/2$,
the dipolar interaction becomes SU(2)-symmetric and conserves the collective spin $S$.
A fully polarized spin state is therefore protected against exchange-induced dephasing even when the pairwise couplings $V_{ij}$ are spatially disordered (Fig.~\ref{fig:floquet}\textbf{d})~\cite{norcia2018cavity,davis2020protecting,lei2025quantum}.

Combining this symmetry protection with a controlled residual anisotropy is predicted to be favorable for spin squeezing near the Heisenberg point~\cite{koyluoglu2025squeezing, kaplan2025theory}.
We implement this control by Floquet engineering the native exchange Hamiltonian $\hat{H}_{\mathrm{XY}}$.
Global $\pi/2$ frame rotations map the interaction into different spin frames, allowing the molecules to evolve under effective Hamiltonians $\hat{H}_{\mathrm{XY}}$, $\hat{H}_{\mathrm{ZY}}$, and $\hat{H}_{\mathrm{XZ}}$ for programmable relative durations $n:m:m$ within each Floquet cycle.
The lowest-order effective Hamiltonian is XXZ
\begin{equation}
    \hat{H}_{\mathrm{XXZ}} =\frac{1}{2}\sum_{i\neq j} \tilde{V}_{ij}
\left(s_i^x s_j^x+s_i^y s_j^y+\Delta s_i^z s_j^z \right)
\end{equation}
with the tunable anisotropy
$\Delta=2m/(m+n)$
and renormalized coupling
$\tilde{V}_{ij}/ V_{ij} =(n+m)/(n+2m).$

We experimentally probe the engineered Hamiltonian by repeating the Ramsey contrast measurement while tuning $\Delta$ from the XY limit, $\Delta=0$, through the isotropic point and towards the Ising side $\Delta=2$.
The many-body dephasing rate $\tilde{\Gamma}$ is strongly reduced as the engineered interaction approaches the SU(2)-symmetric point.
The observed dependence of $\tilde{\Gamma}$ on $\Delta$ is broadly captured by the DTWA simulations using the same independently calibrated experimental parameters.
Small deviations on the Ising side may reflect effects beyond the present DTWA model, including collisional dynamics~\cite{carroll2025observation, wang2025theory, collision_note} and higher-order Floquet corrections.
The simulation further shows that a dominant SU(2)-symmetric interaction can energetically separate collective dynamics from spin-wave excitations, while a small anisotropic component generates spin squeezing in itinerant molecule parity-doublets under realistic experimental conditions~\cite{squeezeing_note}.

Our work transforms parity-doublets from a long-lived single-particle probe to a coherent, interacting, and programmable molecular degree of freedom. 
This is a key step toward generating entangled many-body states within the same manifold used for precision measurements~\cite{bilitewski2021dynamical, wellnitz2024spin}. 
Recent work has proposed that entanglement in parity-doublet states can enhance sensitivity to symmetry-violating moments that are strong probes for new CP-v physics and other new particles, particularly in the low-field regime where parity remains a good quantum number~\cite{zhang2023quantum}.
Future work may extend such entanglement to the polarized, high-electric-field conditions that have complementary relevance for eEDM searches. 
Exploring these systems remains an important open direction.

The same capabilities make interacting polyatomic molecules a new resource for quantum many-body physics.
The system demonstrated here realizes an itinerant dipolar XY spin model and programs a tunable XXZ model through combinations of electric-field control, coherent rf driving, and Floquet protocols~\cite{lu2026probing, gorshkov2011tunable}.
Such tools could generate entangling dynamics beyond the one-axis-twisting interaction studied here, including routes toward programmable spin squeezing and two-axis twisting~\cite{miller2024two, holland2026creating, douglas2025spin}.
Looking forward, loading parity-doublet molecules into dense optical tweezer arrays~\cite{vilas2024optical} would combine long coherence, strong electric-dipole interactions, and the enriched internal structure of polyatomic molecules in a programmable platform for many-body quantum science.


\begin{acknowledgments}
\noindent
{\it Acknowledgments} -- We thank Avikar Periwal for valuable discussions. 
This material is based upon work supported by the Air Force Office of Scientific Research (Award Numbers FA9550-22-1-0228 and FA2386-24-1-4070), the National Science Foundation (PHY-2409404 and NSF Physics Frontier Center PHY-2317134), and the U.S. Army Research Office (W911NF1910283). Support is also acknowledged from the U.S. Department of Energy, Office of Science, National Quantum Information Science Research Centers, Quantum Systems Accelerator. 


\noindent
{\it Author Contributions} -- J.T., P.R., C.H., G.L. designed and performed the experiment, analyzed the data, and conducted numerical simulations. 
J.M.D. oversaw the study. 
All authors discussed and developed an understanding of the results, and contributed to the manuscript.

\noindent
{\it Competing Interests} -- The authors declare that they have no competing financial interests.

\end{acknowledgments}

\bibliography{main}

@PREAMBLE{
 "\providecommand{\noopsort}[1]{}" 
 # "\providecommand{\singleletter}[1]{#1}%" 
}

@article{carr2009cold,
  title={Cold and ultracold molecules: science, technology and applications},
  author={Carr, Lincoln D and DeMille, David and Krems, Roman V and Ye, Jun},
  journal={New Journal of Physics},
  volume={11},
  number={5},
  pages={055049},
  year={2009},
  publisher={IOP Publishing}
}

@article{bohn2017cold,
  title={Cold molecules: Progress in quantum engineering of chemistry and quantum matter},
  author={Bohn, John L and Rey, Ana Maria and Ye, Jun},
  journal={Science},
  volume={357},
  number={6355},
  pages={1002--1010},
  year={2017},
  publisher={American Association for the Advancement of Science}
}

@article{cornish2024quantum,
  title={Quantum computation and quantum simulation with ultracold molecules},
  author={Cornish, Simon L and Tarbutt, Michael R and Hazzard, Kaden RA},
  journal={Nature Physics},
  volume={20},
  number={5},
  pages={730--740},
  year={2024},
  publisher={Nature Publishing Group UK London}
}

@article{yan2013observation,
  title={Observation of dipolar spin-exchange interactions with lattice-confined polar molecules},
  author={Yan, Bo and Moses, Steven A and Gadway, Bryce and Covey, Jacob P and Hazzard, Kaden RA and Rey, Ana Maria and Jin, Deborah S and Ye, Jun},
  journal={Nature},
  volume={501},
  number={7468},
  pages={521--525},
  year={2013},
  publisher={Nature Publishing Group UK London}
}

@article{bao2023dipolar,
  title={Dipolar spin-exchange and entanglement between molecules in an optical tweezer array},
  author={Bao, Yicheng and Yu, Scarlett S and Anderegg, Lo{\"\i}c and Chae, Eunmi and Ketterle, Wolfgang and Ni, Kang-Kuen and Doyle, John M},
  journal={Science},
  volume={382},
  number={6675},
  pages={1138--1143},
  year={2023},
  publisher={American Association for the Advancement of Science}
}

@article{holland2023demand,
  title={On-demand entanglement of molecules in a reconfigurable optical tweezer array},
  author={Holland, Connor M and Lu, Yukai and Cheuk, Lawrence W},
  journal={Science},
  volume={382},
  number={6675},
  pages={1143--1147},
  year={2023},
  publisher={American Association for the Advancement of Science}
}

@article{gregory2024second,
  title={Second-scale rotational coherence and dipolar interactions in a gas of ultracold polar molecules},
  author={Gregory, Philip D and Fernley, Luke M and Tao, Albert Li and Bromley, Sarah L and Stepp, Jonathan and Zhang, Zewen and Kotochigova, Svetlana and Hazzard, Kaden RA and Cornish, Simon L},
  journal={Nature Physics},
  volume={20},
  number={3},
  pages={415--421},
  year={2024},
  publisher={Nature Publishing Group UK London}
}

@article{acme2018improved,
  title={Improved limit on the electric dipole moment of the electron},
  author={{ACME Collaboration}},
  journal={Nature},
  volume={562},
  number={7727},
  pages={355--360},
  year={2018},
  publisher={Nature Publishing Group UK London}
}

@article{kozyryev2017precision,
  title={Precision measurement of time-reversal symmetry violation with laser-cooled polyatomic molecules},
  author={Kozyryev, Ivan and Hutzler, Nicholas R},
  journal={Physical review letters},
  volume={119},
  number={13},
  pages={133002},
  year={2017},
  publisher={APS}
}

@article{anderegg2023quantum,
  title={Quantum control of trapped polyatomic molecules for eEDM searches},
  author={Anderegg, Lo{\"\i}c and Vilas, Nathaniel B and Hallas, Christian and Robichaud, Paige and Jadbabaie, Arian and Doyle, John M and Hutzler, Nicholas R},
  journal={Science},
  volume={382},
  number={6671},
  pages={665--668},
  year={2023},
  publisher={American Association for the Advancement of Science}
}

@article{norrgard2019nuclear,
  title={Nuclear-spin dependent parity violation in optically trapped polyatomic molecules},
  author={Norrgard, Eric B and Barker, Daniel S and Eckel, S and Fedchak, James A and Klimov, Nikoliai N and Scherschligt, Julia},
  journal={Communications Physics},
  volume={2},
  number={1},
  pages={77},
  year={2019},
  publisher={Nature Publishing Group UK London}
}

@article{demille2024quantum,
  title={Quantum sensing and metrology for fundamental physics with molecules},
  author={DeMille, David and Hutzler, Nicholas R and Rey, Ana Maria and Zelevinsky, Tanya},
  journal={Nature Physics},
  volume={20},
  number={5},
  pages={741--749},
  year={2024},
  publisher={Nature Publishing Group UK London}
}

@article{gorshkov2011quantum,
  title={Quantum magnetism with polar alkali-metal dimers},
  author={Gorshkov, Alexey V and Manmana, Salvatore R and Chen, Gang and Demler, Eugene and Lukin, Mikhail D and Rey, Ana Maria},
  journal={Physical Review A—Atomic, Molecular, and Optical Physics},
  volume={84},
  number={3},
  pages={033619},
  year={2011},
  publisher={APS}
}

@article{miller2024two,
  title={Two-axis twisting using Floquet-engineered XYZ spin models with polar molecules},
  author={Miller, Calder and Carroll, Annette N and Lin, Junyu and Hirzler, Henrik and Gao, Haoyang and Zhou, Hengyun and Lukin, Mikhail D and Ye, Jun},
  journal={Nature},
  volume={633},
  number={8029},
  pages={332--337},
  year={2024},
  publisher={Nature Publishing Group UK London}
}

@article{vilas2026quantum,
  title={Quantum-state-controlled collisions of ultracold polyatomic molecules},
  author={Vilas, Nathaniel B and Robichaud, Paige and Hallas, Christian and Tao, Junheng and Anderegg, Lo{\"\i}c and Li, Grace K and Lampson, Hana and Augustovi{\v{c}}ov{\'a}, Lucie D and Bohn, John L and Doyle, John M},
  journal={Physical Review X},
  volume={16},
  number={2},
  pages={021001},
  year={2026},
  publisher={APS}
}

@article{schachenmayer2015many,
  title={Many-body quantum spin dynamics with Monte Carlo trajectories on a discrete phase space},
  author={Schachenmayer, Johannes and Pikovski, Alexander and Rey, Ana Maria},
  journal={Physical Review X},
  volume={5},
  number={1},
  pages={011022},
  year={2015},
  publisher={APS}
}

@article{kitagawa1993squeezed,
  title={Squeezed spin states},
  author={Kitagawa, Masahiro and Ueda, Masahito},
  journal={Physical Review A},
  volume={47},
  number={6},
  pages={5138},
  year={1993},
  publisher={APS}
}

@article{wineland1994squeezed,
  title={Squeezed atomic states and projection noise in spectroscopy},
  author={Wineland, David J and Bollinger, John J and Itano, Wayne M and Heinzen, Daniel J},
  journal={Physical Review A},
  volume={50},
  number={1},
  pages={67},
  year={1994},
  publisher={APS}
}

@article{robichaud2026parity,
  title={Parity-doublet coherence times in optically trapped polyatomic molecules},
  author={Robichaud, Paige and Hallas, Christian and Tao, Junheng and Lee, Giseok and Vilas, Nathaniel B and Doyle, John M},
  journal={Nature},
  volume={651},
  number={8104},
  pages={43--47},
  year={2026},
  publisher={Nature Publishing Group UK London}
}

@article{hallas2023optical,
  title={Optical trapping of a polyatomic molecule in an $\ell$-type parity doublet state},
  author={Hallas, Christian and Vilas, Nathaniel B and Anderegg, Lo{\"\i}c and Robichaud, Paige and Winnicki, Andrew and Zhang, Chaoqun and Cheng, Lan and Doyle, John M},
  journal={Physical Review Letters},
  volume={130},
  number={15},
  pages={153202},
  year={2023},
  publisher={APS}
}

@article{vilas2022magneto,
  title={Magneto-optical trapping and sub-Doppler cooling of a polyatomic molecule},
  author={Vilas, Nathaniel B and Hallas, Christian and Anderegg, Lo{\"\i}c and Robichaud, Paige and Winnicki, Andrew and Mitra, Debayan and Doyle, John M},
  journal={Nature},
  volume={606},
  number={7912},
  pages={70--74},
  year={2022},
  publisher={Nature Publishing Group UK London}
}

@article{kozyryev2017sisyphus,
  title={Sisyphus laser cooling of a polyatomic molecule},
  author={Kozyryev, Ivan and Baum, Louis and Matsuda, Kyle and Augenbraun, Benjamin L and Anderegg, Loic and Sedlack, Alexander P and Doyle, John M},
  journal={Physical review letters},
  volume={118},
  number={17},
  pages={173201},
  year={2017},
  publisher={APS}
}

@article{lasner2025magneto,
  title={Magneto-optical trapping of a heavy polyatomic molecule for precision measurement},
  author={Lasner, Zack D and Frenett, Alexander and Sawaoka, Hiromitsu and Anderegg, Lo{\"\i}c and Augenbraun, Benjamin and Lampson, Hana and Li, Mingda and Lunstad, Annika and Mango, Jack and Nasir, Abdullah and others},
  journal={Physical Review Letters},
  volume={134},
  number={8},
  pages={083401},
  year={2025},
  publisher={APS}
}

@article{hallas2026high,
  title={High compression blue-detuned magneto-optical trap of polyatomic molecules},
  author={Hallas, Christian and Li, Grace K and Vilas, Nathaniel B and Robichaud, Paige and Anderegg, Lo{\"\i}c and Doyle, John M},
  journal={Physical Review Letters},
  volume={136},
  number={13},
  pages={133402},
  year={2026},
  publisher={APS}
}

@article{li2025conveyor,
  title={Conveyor-belt magneto-optical trapping of molecules},
  author={Li, Grace K and Hallas, Christian and Doyle, John M},
  journal={New Journal of Physics},
  volume={27},
  number={4},
  pages={043002},
  year={2025},
  publisher={IOP Publishing}
}

@article{zhang2023quantum,
  title={Quantum-enhanced metrology for molecular symmetry violation using decoherence-free subspaces},
  author={Zhang, Chi and Yu, Phelan and Jadbabaie, Arian and Hutzler, Nicholas R},
  journal={Physical Review Letters},
  volume={131},
  number={19},
  pages={193602},
  year={2023},
  publisher={APS}
}

@article{holland2026creating,
  title={Creating and Probing Spin-Squeezed States of Molecules},
  author={Holland, Connor M and Welsh, Callum L and Lu, Yukai and Wellnitz, David and Chen, Xing-Yan and Rey, Ana Maria and Cheuk, Lawrence W},
  journal={arXiv preprint arXiv:2606.02500},
  year={2026}
}

@article{lu2026probing,
  title={Probing Coherent Many-Body Spin Dynamics in a Molecular Tweezer Array Quantum Simulator},
  author={Lu, Yukai and Holland, Connor M and Welsh, Callum L and Chen, Xing-Yan and Cheuk, Lawrence W},
  journal={arXiv preprint arXiv:2603.19090},
  year={2026}
}

@article{vilas2024optical,
  title={An optical tweezer array of ultracold polyatomic molecules},
  author={Vilas, Nathaniel B and Robichaud, Paige and Hallas, Christian and Li, Grace K and Anderegg, Lo{\"\i}c and Doyle, John M},
  journal={Nature},
  volume={628},
  number={8007},
  pages={282--286},
  year={2024},
  publisher={Nature Publishing Group UK London}
}

@article{schachenmayer2015dynamics,
  title={Dynamics of correlations in two-dimensional quantum spin models with long-range interactions: a phase-space Monte-Carlo study},
  author={Schachenmayer, Johannes and Pikovski, Alexander and Rey, Ana Maria},
  journal={New Journal of Physics},
  volume={17},
  number={6},
  pages={065009},
  year={2015},
  publisher={IOP Publishing}
}

@article{zhu2019generalized,
  title={A generalized phase space approach for solving quantum spin dynamics},
  author={Zhu, Bihui and Rey, Ana Maria and Schachenmayer, Johannes},
  journal={New Journal of Physics},
  volume={21},
  number={8},
  pages={082001},
  year={2019},
  publisher={IOP Publishing}
}

@article{gullion1990new,
  title={New, compensated carr-purcell sequences},
  author={Gullion, Terry and Baker, David B and Conradi, Mark S},
  journal={Journal of Magnetic Resonance (1969)},
  volume={89},
  number={3},
  pages={479--484},
  year={1990},
  publisher={Elsevier}
}

@article{pezze2018quantum,
  title={Quantum metrology with nonclassical states of atomic ensembles},
  author={Pezze, Luca and Smerzi, Augusto and Oberthaler, Markus K and Schmied, Roman and Treutlein, Philipp},
  journal={Reviews of Modern Physics},
  volume={90},
  number={3},
  pages={035005},
  year={2018},
  publisher={APS}
}

@article{li2023tunable,
  title={Tunable itinerant spin dynamics with polar molecules},
  author={Li, Jun-Ru and Matsuda, Kyle and Miller, Calder and Carroll, Annette N and Tobias, William G and Higgins, Jacob S and Ye, Jun},
  journal={Nature},
  volume={614},
  number={7946},
  pages={70--74},
  year={2023},
  publisher={Nature Publishing Group UK London}
}

@article{christakis2023probing,
  title={Probing site-resolved correlations in a spin system of ultracold molecules},
  author={Christakis, Lysander and Rosenberg, Jason S and Raj, Ravin and Chi, Sungjae and Morningstar, Alan and Huse, David A and Yan, Zoe Z and Bakr, Waseem S},
  journal={Nature},
  volume={614},
  number={7946},
  pages={64--69},
  year={2023},
  publisher={Nature Publishing Group UK London}
}

@article{micheli2006toolbox,
  title={A toolbox for lattice-spin models with polar molecules},
  author={Micheli, Andrea and Brennen, Gavin K and Zoller, Peter},
  journal={Nature Physics},
  volume={2},
  number={5},
  pages={341--347},
  year={2006},
  publisher={Nature Publishing Group UK London}
}

@article{gorshkov2011tunable,
  title={Tunable superfluidity and quantum magnetism with ultracold polar molecules},
  author={Gorshkov, Alexey V and Manmana, Salvatore R and Chen, Gang and Ye, Jun and Demler, Eugene and Lukin, Mikhail D and Rey, Ana Maria},
  journal={Physical review letters},
  volume={107},
  number={11},
  pages={115301},
  year={2011},
  publisher={APS}
}

@article{roussy2023improved,
  title={An improved bound on the electron’s electric dipole moment},
  author={Roussy, Tanya S and Caldwell, Luke and Wright, Trevor and Cairncross, William B and Shagam, Yuval and Ng, Kia Boon and Schlossberger, Noah and Park, Sun Yool and Wang, Anzhou and Ye, Jun and others},
  journal={Science},
  volume={381},
  number={6653},
  pages={46--50},
  year={2023},
  publisher={American Association for the Advancement of Science}
}

@article{picard2025entanglement,
  title={Entanglement and iSWAP gate between molecular qubits},
  author={Picard, Lewis RB and Park, Annie J and Patenotte, Gabriel E and Gebretsadkan, Samuel and Wellnitz, David and Rey, Ana Maria and Ni, Kang-Kuen},
  journal={Nature},
  volume={637},
  number={8047},
  pages={821--826},
  year={2025},
  publisher={Nature Publishing Group UK London}
}

@article{ruttley2025long,
  title={Long-lived entanglement of molecules in magic-wavelength optical tweezers},
  author={Ruttley, Daniel K and Hepworth, Tom R and Guttridge, Alexander and Cornish, Simon L},
  journal={Nature},
  volume={637},
  number={8047},
  pages={827--832},
  year={2025},
  publisher={Nature Publishing Group UK London}
}

@article{demille2002quantum,
  title={Quantum computation with trapped polar molecules},
  author={DeMille, David},
  journal={Physical Review Letters},
  volume={88},
  number={6},
  pages={067901},
  year={2002},
  publisher={APS}
}

@article{bilitewski2021dynamical,
  title={Dynamical generation of spin squeezing in ultracold dipolar molecules},
  author={Bilitewski, Thomas and De Marco, Luigi and Li, Jun-Ru and Matsuda, Kyle and Tobias, William G and Valtolina, Giacomo and Ye, Jun and Rey, Ana Maria},
  journal={Physical Review Letters},
  volume={126},
  number={11},
  pages={113401},
  year={2021},
  publisher={APS}
}

@article{wellnitz2024spin,
  title={Spin squeezing with itinerant dipoles: A case for shallow lattices},
  author={Wellnitz, David and Mamaev, Mikhail and Bilitewski, Thomas and Rey, Ana Maria},
  journal={Physical Review Research},
  volume={6},
  number={1},
  pages={L012025},
  year={2024},
  publisher={APS}
}

@article{vilas2023blackbody,
  title={Blackbody thermalization and vibrational lifetimes of trapped polyatomic molecules},
  author={Vilas, Nathaniel B and Hallas, Christian and Anderegg, Lo{\"\i}c and Robichaud, Paige and Zhang, Chaoqun and Dawley, Sam and Cheng, Lan and Doyle, John M},
  journal={Physical Review A},
  volume={107},
  number={6},
  pages={062802},
  year={2023},
  publisher={APS}
}

@article{koyluoglu2025squeezing,
  title={Squeezing towards the Heisenberg limit with locally interacting spins},
  author={Koyluoglu, Nazli Ugur and Rajagopal, Shankari V and Moreau, Gabriel L and Hines, Jacob A and Markovi{\'c}, Ognjen and Schleier-Smith, Monika},
  journal={arXiv preprint arXiv:2506.16973},
  year={2025}
}

@article{takahashi2026engineered,
  title={Engineered molecular clock transitions for precision measurements},
  author={Takahashi, Yuiki and Ramachandran, Harish D and Jadbabaie, Arian and Zeng, Yi and Zhang, Chi and Hutzler, Nicholas R},
  journal={Physical Review X},
  volume={16},
  number={3},
  pages={031011},
  year={2026},
  publisher={APS}
}

@article{zhou2020quantum,
  title={Quantum metrology with strongly interacting spin systems},
  author={Zhou, Hengyun and Choi, Joonhee and Choi, Soonwon and Landig, Renate and Douglas, Alexander M and Isoya, Junichi and Jelezko, Fedor and Onoda, Shinobu and Sumiya, Hitoshi and Cappellaro, Paola and others},
  journal={Physical review X},
  volume={10},
  number={3},
  pages={031003},
  year={2020},
  publisher={APS}
}

@misc{collision_note,
  note = {The DTWA simulation describes the long-range interaction-induced phase during molecular encounters, but neglects momentum backaction at short range and spin-motion entanglement. We present an additional two-body scattering S-matrix calculation in the SI.},
}

@article{carroll2025observation,
  title={Observation of generalized tJ spin dynamics with tunable dipolar interactions},
  author={Carroll, Annette N and Hirzler, Henrik and Miller, Calder and Wellnitz, David and Muleady, Sean R and Lin, Junyu and Zamarski, Krzysztof P and Wang, Reuben RW and Bohn, John L and Rey, Ana Maria and others},
  journal={Science},
  volume={388},
  number={6745},
  pages={381--386},
  year={2025},
  publisher={American Association for the Advancement of Science}
}

@misc{squeezeing_note,
  note = {After generating squeezed states via the OAT dynamics, one can Floquet engineer to $\Delta=1$ to protect the state from dephasing during the interrogation. 
  Another option is to drive simultaneous transitions to $m_F=0$ in the same parity doublet manifold, where the transition dipole moment and the interactions vanish.
  },
}

@article{norcia2018cavity,
  title={Cavity-mediated collective spin-exchange interactions in a strontium superradiant laser},
  author={Norcia, Matthew A and Lewis-Swan, Robert J and Cline, Julia RK and Zhu, Bihui and Rey, Ana M and Thompson, James K},
  journal={Science},
  volume={361},
  number={6399},
  pages={259--262},
  year={2018},
  publisher={American Association for the Advancement of Science}
}

@article{davis2020protecting,
  title={Protecting spin coherence in a tunable Heisenberg model},
  author={Davis, Emily J and Periwal, Avikar and Cooper, Eric S and Bentsen, Gregory and Evered, Simon J and Van Kirk, Katherine and Schleier-Smith, Monika H},
  journal={Physical Review Letters},
  volume={125},
  number={6},
  pages={060402},
  year={2020},
  publisher={APS}
}

@article{nasir2026order,
  title={Order of Magnitude Improved Optical Trapping of Molecules Through Transverse Cooling},
  author={Nasir, Abdullah and Lunstad, Annika and Li, Mingda and Salim, Saif and Liu, Shuqi and Hallas, Christian and Lasner, Zack and Doyle, John M},
  journal={arXiv preprint arXiv:2607.16913},
  year={2026}
}

@article{wang2025theory,
  title={Theory of itinerant collisional spin dynamics in nondegenerate molecular gases},
  author={Wang, Reuben RW and Bohn, John L},
  journal={Physical Review A},
  volume={112},
  number={4},
  pages={043315},
  year={2025},
  publisher={APS}
}

@article{douglas2025spin,
  title={Spin Squeezing with Itinerant Magnetic Dipoles},
  author={Douglas, Alec and Kaxiras, Vassilios and Su, Lin and Szurek, Michal and Singh, Vikram and Markovi{\'c}, Ognjen and Greiner, Markus},
  journal={Physical Review X},
  volume={15},
  number={4},
  pages={041021},
  year={2025},
  publisher={APS}
}

@article{choi2020robust,
  title={Robust dynamic Hamiltonian engineering of many-body spin systems},
  author={Choi, Joonhee and Zhou, Hengyun and Knowles, Helena S and Landig, Renate and Choi, Soonwon and Lukin, Mikhail D},
  journal={Physical Review X},
  volume={10},
  number={3},
  pages={031002},
  year={2020},
  publisher={APS}
}

@article{idziaszek2010universal,
  title={Universal rate constants for reactive collisions of ultracold molecules},
  author={Idziaszek, Zbigniew and Julienne, Paul S},
  journal={Physical review letters},
  volume={104},
  number={11},
  pages={113202},
  year={2010},
  publisher={APS}
}

@article{kaplan2025theory,
  title={Theory of Scalable Spin Squeezing with Disordered Quantum Dipoles},
  author={Kaplan-Lipkin, Avi and Crowley, Philip JD and Hall{\'e}n, Jonathan N and Wang, Zilin and Wu, Weijie and Chern, Sabrina and Laumann, Chris R and Pollet, Lode and Yao, Norman Y},
  journal={arXiv preprint arXiv:2512.19781},
  year={2025}
}

@article{milner2025coherent,
  title={Coherent evolution of superexchange interaction in seconds-long optical clock spectroscopy},
  author={Milner, William R and Lannig, Stefan and Mamaev, Mikhail and Yan, Lingfeng and Chu, Anjun and Lewis, Ben and Frankel, Max N and Hutson, Ross B and Rey, Ana Maria and Ye, Jun},
  journal={Science},
  volume={388},
  number={6746},
  pages={503--508},
  year={2025},
  publisher={American Association for the Advancement of Science}
}

@article{lei2025quantum,
  title={Quantum thermalization and Floquet engineering in a spin ensemble with a clock transition},
  author={Lei, Mi and Fukumori, Rikuto and Wu, Chun-Ju and Barnes, Edwin and Economou, Sophia E and Choi, Joonhee and Faraon, Andrei},
  journal={Nature Physics},
  volume={21},
  number={8},
  pages={1196--1202},
  year={2025},
  publisher={Nature Publishing Group UK London}
}

\clearpage

\appendix


\setcounter{figure}{0}
\renewcommand{\thefigure}{S\arabic{figure}}

\renewcommand{\theHfigure}{S\arabic{figure}}

\section*{Experimental sequence}

The experiment prepares ultracold CaOH molecules in the parity-doublet manifold using a similar sequence as detailed in the previous works~\cite{vilas2022magneto, robichaud2026parity, anderegg2023quantum}. 
After laser cooling and trapping molecules in the blue-detuned conveyor-belt MOT, we turn on the cODT while applying the single-frequency grey molasses cooling for $15~\mathrm{ms}$, which cools the molecules to $60~\mathrm{\mu K}$ in the trap.
Here, we take a first fluorescence image of the trapped cloud for $7~\mathrm{ms}$ to provide a reference for future number normalization and data thresholding.
These molecules are in the vibrational ground state $\tilde{X}(000)$ and are recooled by the same lasers shortly after the image.
Next, we optically pump them into the vibrational bending mode $\tilde{X}(010)$ via the excited state $\tilde{A}(010)$.
As there are four parity-allowed hyperfine states into which $\tilde{A}(010)$ spontaneously decays, we use frequency-modulated microwave to selectively address all but $\ket{N=1, J=1/2, F=0, p=-}$.
By transferring the other three states to $N=2$ and further to $\tilde{A}(010)$ with an $N=2$ repump laser, we are able to prepare molecules to this single state. 
We calibrate the state preparation efficiency from $\tilde{X}(000)$ to $\ket{\uparrow}$ is $33(2)\%$.

From $\ket{N=1, J=1/2, F=0, p=-}$ to $\ket{\uparrow}\equiv\ket{N=1,J=3/2,F=2,m_F=2,p=+}$, We drive an rf adiabatic sweep at $72.4~\mathrm{MHz}$ by ramping the bias $B_z$ from $1.2$ to $2.7~\mathrm{G}$ within $12~\mathrm{ms}$.
A resonant pushout laser pulse is applied for $3~\mathrm{ms}$ to clear all remaining $\ket{p=-}$ states, and the bias $B_z$ is ramped up to about $30~\mathrm{G}$.
Here, we apply the rf Ramsey plus dynamical decoupling pulses.
To do this, the rf source is switched to a high-speed arbitrary waveform generator during the interrogation time $T$.
Finally, we deplete the $\ket{p=-}$ states by another push-out pulse, and time-reverse the rf adiabatic sweep to only probe the population in the $\ket{\uparrow}$ state using the same single-frequency cooling light for $30~\mathrm{ms}$.

\section*{Crossed ODT calibrations}

Our crossed optical dipole trap (cODT) is made by two tightly focused $\lambda=1064~\mathrm{nm}$ laser beams intersecting in the $xy$ plane.
The two beams are produced by independent lasers and therefore are non-interfering.
Beam 1 transmits through a telescope that only expands the beam in $z$ by about 5x in the Fourier plane and is further focused into an elliptical beam with $1/e^2$ waists $(w_y, w_z) = (50, 11)~\mathrm{\mu m}$ at the molecular plane.
Beam 2 is expanded by an isotropic telescope and focused to $(w_{x-y}, w_z) = (26, 26)~\mathrm{\mu m}$.
Beam 2 is combined with the $x-y$ MOT beam on a dichroic mirror and sent into the chamber. 
Both dipole beams have linear polarizations controlled by two motorized half-wave plates.

We calibrate the radial harmonic trap frequencies of each trap separately using parametric heating. 
In this approach, we perturbatively modulate the trap intensity, resulting in heating near twice the trap frequencies at each principal trap axis.
This temperature change is converted to molecule loss by deliberately snapping the trap off for $\sim 150~\mathrm{\mu s}$ and then turning it on to recapture. 
For Beam 1, we measure the radial frequencies to be $(\omega_{1,y}, \omega_{1, z})/2\pi= (2.2(1), 11.5(2))~\mathrm{kHz}$ at total power $P_1=16.5~\mathrm{W}$.
For Beam 2, the radial frequencies are $(\omega_{2, x-y}, \omega_{2, z})/2\pi= (3.0(1), 3.0(1))~\mathrm{kHz}$ at total power $P_2= 15~\mathrm{W}$.

Because the two beams intersect at $45\mathrm{^\circ}$, the trap frequencies hybridize in the $xy$ plane, and add in $z$:
\begin{align}
\quad \omega_{\pm}^2 =& \frac{1}{2}\left[\omega_{1,y}^2+\omega_{2,x-y}^2\pm\sqrt{\omega_{1,y}^4+\omega_{2,x-y}^4}\right], \\
\quad \omega_z^2= &\omega_{1,z}^2+\omega_{2,z}^2.
\end{align}
Plugging in individual frequencies, we expect $(\omega_{-}, \omega_{+}, \omega_z)/2\pi= (1.35(5), 3.47(9), 11.8(2))~\mathrm{kHz}$.
To verify this, we modulate the two beam intensities simultaneously in phase in the cODT to measure $\omega/2\pi= (1.3(1), 3.5(1), 11.5(2))~\mathrm{kHz}$. 

The two dipole beams traverse distinct optical paths and we find it necessary to frequently check the focuses are well overlapped in space.
For this positional fine-tuning, we include a piezo-driven mirror in each beam's path.
We use three methods for the overlap: (i) the initial alignment of the ODTs is individually optimized for the number of loaded molecules from the blue-detuned conveyor-belt MOT, which has a $1/e^2$ radius $\sim 60~\mathrm{\mu m}$;
(ii) a ``hand-off'' procedure, where the overlap condition maximizes the fraction transferred adiabatically from Trap 1 to Trap 2 and back;
(iii) trap frequencies are measured as we scan the beam positions.
All three methods agree on locating the spatial overlap within $2~\mathrm{\mu m}$ precision.

\section*{Molecular density calibration}

We independently calibrate the trapped molecular density in two ways:
(i) measuring the molecule number and temperature separately and
combining them with the calibrated trap geometry, and (ii) using the
measured two-body inelastic loss rate together with the collisional rate
constant reported in Ref.~\cite{vilas2026quantum}.

\emph{(i) Number, temperature, and trap geometry.}---For a thermal
cloud in the harmonic trap, the peak number density is
$
n_0=N\,\omega_x\omega_y\omega_z
\left({m}/{2\pi k_B T}\right)^{3/2}
$. Therefore, the
calibration reduces to determining $N$, $T$, and the trap frequencies
$\omega_\mu$.

The molecule number is obtained from fluorescence imaging. 
We calibrate the imaging collection efficiency $\eta_{coll}=1.3\%$ by comparing the counts collected from a known number of photons, using an imaging system with effective $\mathrm{NA}=0.3$ similar to that of Refs.~\cite{robichaud2026parity, vilas2026quantum}. 
We then calibrate the photon scattering rate to be $22~\mathrm{kHz}$ in the same imaging condition.
Next, we separately measure the EMCCD counts per incident photon to be $114$.
These combined with the state preparation efficiency, calculate the molecule number $N$ in the parity-doublet manifold.
We estimate a combined fractional uncertainty of 25\% in $N$, which dominates the density uncertainty.

The temperature is measured by a trap-depth ``shaving'': we adiabatically lower the trap depth to a variable value $U$, losing molecules with energy exceeding $U$, and record the surviving fraction. 
Fitting the survival curve to a thermal energy distribution yields $T=60(8)~\mu$K. 
This method is well suited to the small molecule numbers here, where time-of-flight imaging has limited signal-to-noise.

The trap frequencies are measured by parametric heating spectroscopy.
Because the trap depth greatly exceeds the temperature $\eta=U_0/k_B T=22(1)$, the molecules occupy only the harmonic bottom of the trap and anharmonic corrections are small.

\emph{(ii) Two-body loss.}---As an independent check, we use the inelastic collisional loss of molecules confined at high density. 
The loss rate constant $k$ for molecules occupying states in $N=1$ of the $\tilde X(010)$ bending mode was measured in Ref.~\cite{vilas2026quantum}.
Recording the molecule number versus hold time and fitting to the two-body rate equation $\dot N=-k\langle n\rangle N - N/\tau$, where $\langle n\rangle=n_0/2^{3/2}$ is the average density and $\tau$ the single-molecule lifetime, we invert the observed decay to obtain a density.
We specifically choose $|\!\downarrow\rangle=|N=1,J=3/2,F=2,m_F=2,p=-\rangle$ for this measurement for its higher collisional rate.
The two methods agree within their combined uncertainty, which we take as the 25\% error bar on $n_0$ in the parity-doublet manifold.
When we need to scan $n_0$, we vary the laser-slowing time to change the initial number loaded in the cODT.

\section*{Single-particle coherence time}

Observing dipolar interactions requires a suppression of the single-body decoherence rate.
The parity-doublet transition is sensitive to the electric field environment as well as a parity-dependent trap shift.
Since the former shift is quadratic, we use in-vacuum electrodes and coils to apply dc electric fields that cancel the ambient lab field. 
We also bathe our vacuum chamber with UV LEDs to desorb charges introduced by the buffer gas beam to increase stability.

To reach high enough density to observe dipolar interaction, we run the experiment at a trap depth about $35\times$ higher than Ref.~\cite{robichaud2026parity}.
Owing to the $U_0^{3/2}$ scaling of the decoherence rate, the differential trap shifts are the dominant limitation to the single-spin coherence time.
We conduct iterations of optimization on the magic condition of the trap polarizations.
We report an optimal bare coherence time of $T_2=2.5(4)~\mathrm{ms}$ in the cODT at $U_0/k_B=1.3~\mathrm{mK}$.

\section*{Dynamical decoupling sequences}

The timescale set by dipolar interactions is $\sim 50\mathrm{ms}$, and dynamical decoupling is essential for extending the single-spin coherence time beyond it.
In particular, we use pulse sequences from the XY family, in which nominal $\pi$ pulses are applied about alternating axes in the equatorial plane.
Ideally, these pulse sequences create a bandpass filter that rejects detuning noises with frequencies other than near $1/2\tau_p$, while leaving the interaction Hamiltonian intact.
Compared with Carr--Purcell sequences that use pulses about a single axis, XY sequences cancel pulse imperfections more symmetrically.

In the cODT, because the molecules undergo harmonic motion in the trap, the dominant single-particle dephasing arises from the residual parity-dependent AC Stark shift modulated primarily at the trap frequencies and their harmonics.
The dominant spectral weight of this detuning noise is therefore bounded from below by approximately twice the largest trap frequency $\sim 23~\mathrm{kHz}$.
We choose the interpulse spacing $\tau_p = 25~\mu\mathrm{s}$.
The pulse duration is scanned and chosen so that its Fourier broadening does not drive nearby hyperfine transitions.

For typical Ramsey measurements with $T = 50$--$100~\mathrm{ms}$, the sequence contains $2.5$--$5\times 10^3$ refocusing pulses.
Even though the XY family pulses can self-correct, this large number of pulses makes robustness to pulse imperfections a stringent requirement for preserving consistent Ramsey contrast and phase.
In the experiment, we deliberately scan the pulse duration and observe typical pulse error tolerance at $\sim 1\%$ of the pulse area.
We reproduce this behavior by Monte Carlo simulations that include detuning noise under experimental parameters and model XY pulse errors by imperfect rotations. 
Such imperfections are likely from slow drifts and fluctuations of the rf amplifier gain associated with thermal relaxation.

We achieve the requirement using three methods simultaneously.
We first calibrate the $\pi$ pulse area by preparing $\ket{\uparrow}$ and applying $\sim 50$ pairs of $X$ pulses, where an ideal pulse pair should preserve the initial spin.
During the experiment run, we monitor the rf intensity in real time.
Finally, we upgrade the sequence to use XY16 ($X\,Y\,X\,Y\,Y\,X\,Y\,X\,Y\,X\,Y\,X\,X\,Y\,X\,Y$), which suppresses the pulse imperfection effects to higher order.

\begin{figure}
\centering
\includegraphics{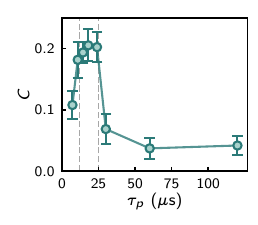}
\caption{Unormalized Ramsey contrast at $T=75~\mathrm{ms}$ as function of the dynamical decoupling pulse interval $\tau_p$.
Dashed vertical lines are $\tau_p=12$ and $25~\mathrm{\mu s}$.
Error bars represent $68\%$ confidence intervals from bootstrapping.
}
\label{fig:tau_scan}
\end{figure}

\section*{Floquet-engineered interactions}

To program the interaction anisotropy, we periodically interleave the free interaction evolution with global $\pi/2$ frame-rotation pulses, following the average-Hamiltonian approach used for dipolar spin systems~\cite{zhou2020quantum, miller2024two, choi2020robust,lei2025quantum}.
Within one Floquet cycle $\tau_c$, the spins evolve in the toggling frame set by the applied pulses, under the interaction Hamiltonian expressed in different spin frames.
A $\pi/2$ rotation about $x$ sends $y\to z$ and leaves $x$ fixed, mapping the interaction to $\widehat{XX}+\widehat{ZZ}\equiv\hat{H}_{XZ}$.
Cycling through these frames therefore lets the molecules evolve under $\hat{H}_{XY}$, $\hat{H}_{YZ}$, and $\hat{H}_{XZ}$ for programmable relative durations.
The exact ordering of frame-switching pulses and interaction intervals is shown in Fig.~4\textbf{a}.
 
We allocate durations $n:m:m$ to the $XY$, $YZ$, and $XZ$ frames within one Floquet cycle of period $\tau_c\propto n+2m$. Provided the cycle is short compared to the interaction timescale, the zeroth-order Magnus expansion is the time average over the cycle,
\begin{equation}
  \hat{H}^{(0)}
  = \frac{1}{n+2m}\left[\,n\,\hat{H}_{\mathrm{XY}}
      + m\,\hat{H}_{\mathrm{YZ}} + m\,\hat{H}_{\mathrm{XZ}}\right].
  \label{eq:Havg}
\end{equation}
Substituting the frame-transformed interactions and collecting terms,
\begin{equation}
  \hat{H}_{\mathrm{XXZ}}
  = \frac{n+m}{n+2m}\,\frac{1}{2}\sum_{i\neq j} V_{ij}
    \left(s^{x}_{i}s^{x}_{j}+s^{y}_{i}s^{y}_{j}
          +\Delta\,s^{z}_{i}s^{z}_{j}\right)
\end{equation}
with 
$\Delta=2m/(m+n)$.
The prefactor $(n+m)/(n+2m)$ is an overall renormalization of the interaction strength.
Within each interaction interval, we embed XY16 $\pi$-pulse trains with spacing $\tau_p$ to refocus single-particle dephasing.
For the experiment, we typically select $\tau_c=190~\mathrm{\mu s}$ and $\tau_p=12~\mathrm{\mu s}$.

By calibrating using Ramsey, we experimentally identify that the frame-switching $\pi/2$ rotations are the dominant source of coherent pulse error in the sequence. 
We implement them as BB1 composite pulses to suppress sensitivity to rf pulse-area
fluctuations. In addition, we alternate two Floquet cycles, denoted $(+)$ and $(-)$, that apply the frame rotations in opposite senses (Fig.~4\textbf{a}). The leading-order coherent error accumulated from the frame-switching pulses reverses sign between the two cycles, in analogy to the cancellation by the XY family.

\begin{figure}
\centering
\includegraphics{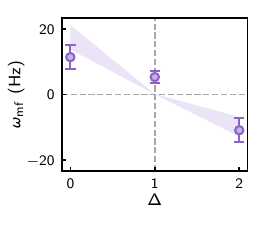}
\caption{Mean-field spin precession rate as a function of anisotropy $\Delta$ in the Floquet-engineered Hamiltonian.
The peak density $n_0=3.0\times 10^{10}~\mathrm{cm^{-3}}$.
Error bars represent $68\%$ confidence intervals from bootstrapping.
Shaded region represents the DTWA simulation prediction.
}
\label{fig:floquet_mf}
\end{figure}

In the experiment, we perform similar Ramsey measurements to determine the dephasing rate $\Gamma$.
To ensure that complex Floquet-engineering pulses do not introduce decoherence that mixes with many-body dephasing, we separately measure the decoherence rate $\Gamma_{\mathrm{low}}$ at $\sim 10\%$ the maximum density for each $\Delta$, and subtract it $\tilde \Gamma = \Gamma-\Gamma_{\mathrm{low}}$.
To verify that the engineered Hamiltonian realizes the intended $\Delta$, we probe the spin-precession dynamics using the same protocol as in Fig.~\ref{fig:precession}.
The observed suppression and subsequent reversal of the precession direction with varying $\Delta$ in Fig.~\ref{fig:floquet_mf} demonstrate that the system is tuned through the isotropic Heisenberg point.

The leading next-order correction from the Magnus expansion is
$\sim\tau_c\,[\hat{H}_{\mathrm{frame}},\hat{H}_{\mathrm{frame}'}]$ between the interaction in different frames, which are suppressed linearly in $\tau_c$.
Residual deviations from the DTWA prediction observed on the Ising side of Fig.~4\textbf{b} may additionally reflect these higher-order Floquet corrections.

\section*{DTWA simulation}

We model the many-body spin dynamics with the discrete truncated
Wigner approximation (DTWA)~\cite{schachenmayer2015dynamics,
schachenmayer2015many}.
This method captures the leading quantum fluctuations of the collective spin by sampling the initial state over a discrete phase-space distribution and propagating each sample under the classical equations of motion.
Because the molecules move during the interaction time, each trajectory carries its own set of time-dependent dipolar couplings generated by the thermal motion in the trap.

Each molecule is a pseudo-spin-$1/2$, represented by a classical vector $\mathbf{s}_i=(s^x_i,s^y_i,s^z_i)$.
For an initial coherent state with mean spin along the direction $(\theta,\phi)$, we sample from the spin-$1/2$ discrete Wigner distribution in a frame rotated by $R^T(\theta,\phi)$. 
Each sampled vector is normalized to length $\sqrt{3}/2$.

For each realization we draw the initial positions $\mathbf{r}_i(0)$ and velocities $\mathbf{v}_i(0)$ of the $N$ molecules from the Maxwell-Boltzmann distribution of the harmonic trap at temperature $T$.
As an approximation that neglects the backaction force, the evolution the molecules follows independent single-particle harmonic trajectories,
$
r_{i,\mu}(t)=r_{i,\mu}(0)\cos(\omega_\mu t)
+v_{i,\mu}(0)/{\omega_\mu}\sin(\omega_\mu t),
$
where $\mu=x,y,z$.
At each time step, the pairwise couplings are evaluated from the instantaneous positions.

Each classical spin precesses in the mean field
of all the others, 
$
\dot{\mathbf{s}}_i=\mathbf{B}_i\times\mathbf{s}_i,
$
where 
$
\mathbf{B}_i=\sum_{j\neq i}V_{ij}(t)\,(s^x_j,\,s^y_j,\,0)/\hbar.
$
We integrate these with a symmetric operator-splitting (Trotter) scheme built from exact two-spin propagators. 
For each pair, the exchange propagator is Strang-split into rotations generated by the $s^x s^x$ and $s^y s^y$ couplings
\begin{equation}
e^{dt\left(\mathcal{L}_{ij}^{(x)}+\mathcal{L}_{ij}^{(y)}\right)}
=
e^{\frac{dt}{2}\mathcal{L}_{ij}^{(x)}}
e^{dt\mathcal{L}_{ij}^{(y)}}
e^{\frac{dt}{2}\mathcal{L}_{ij}^{(x)}}
+\mathcal{O}\!\left(dt^3\right),
\end{equation}
where
$
\mathcal{L}_{ij}^{(x)}= V_{ij}s_i^x s_j^x,\ \mathcal{L}_{ij}^{(y)} =  V_{ij}s_i^y s_j^y.
$
Furthermore, the full time step is symmetrized by sweeping the pair indices in ascending order first and descending
\begin{equation}
e^{\frac{dt}{2}\mathcal{L}_1}
\cdots
e^{\frac{dt}{2}\mathcal{L}_{M-1}}
e^{dt\mathcal{L}_M}
e^{\frac{dt}{2}\mathcal{L}_{M-1}}
\cdots
e^{\frac{dt}{2}\mathcal{L}_1},
\end{equation}
where $\mathcal{L}_p$ spans all the pairs.
The scheme is accurate to order $\mathcal{O}(dt^3)$ for a single update.
This is essential because the thermal motion produces couplings spanning several orders of magnitude, for which an explicit Euler method requires a prohibitively small time step.

For each set of parameters, we average over $N_{\rm traj}$ Wigner
spin samples and $N_{\rm init}$ independent motional realizations.
The collective spin components $S_\alpha=\sum_i s^\alpha_i$ yield the transverse contrast $S_\perp(T)/S_\perp(0)$ used in the dephasing measurement (Fig.~2, 4) and the azimuthal precession angle $\varphi(T)=\arg(S_x+iS_y)$ used in the mean-field precession measurement (Fig.~3).
For these simulation data, we select $N_{\rm traj}=60$, $N_{\rm init}=30$ and $dt=2.5~\mathrm{\mu s}$.

\section*{Collisional framework}

The DTWA simulation captures coherent, long-range spin evolution, including the phase accumulated during repeated encounters, but neglects interaction-induced changes of the trajectories, spin-motion entanglement, and spin-flipping collisions.
We therefore treat DTWA as the many-body model for coherent long-range spin evolution and independently assess two-body scattering, as detailed in the SI.

For the scattering calculation, an incoming pair channel is denoted by
\begin{equation}
q\in
\left\{
|\uparrow\uparrow\rangle,
|T_0\rangle,
|\downarrow\downarrow\rangle
\right\},
\qquad
|T_0\rangle=
\frac{
|\uparrow\downarrow\rangle+
|\downarrow\uparrow\rangle
}{\sqrt{2}}.
\end{equation}
For a common relative collision energy
$
E_{\rm c}={\hbar^2k^2}/{2\mu},
$
the branch with asymptotic internal energy
$
\epsilon_q
$
is propagated at total energy
$
E_q=\epsilon_q+E_{\rm c}.
$
Amplitudes from different incoming spin components are added coherently when they lead to the same outgoing kinetic energy and orbital channel.
Tracing over the outgoing motional state then gives the scattering generator for the internal pair state: 
scattering that is identical for all spin components leaves the Ramsey coherence unchanged, whereas spin-dependent scattering into distinguishable motional channels reduces it.
Missing outgoing flux is identified with short-range universal loss. 
The physical scattering framework follow Ref.~\cite{vilas2026quantum} but are reduced to the parity-doublet manifold: we solve three-dimensional coupled radial equations in an exchange-symmetrized bosonic basis and impose a universal absorptive boundary at short range.
We average the resulting rate coefficients over the thermal distribution and the density profile of the cloud.

The collisional calculation reproduces a similar Ramsey dephasing rate and density dependence to DTWA.
Physically, the scattering and DTWA frameworks both contain contributions from the same long-range tail of the dipolar interaction; their dephasing rates are not additive. 
In our regime $C_3 n_0/k_B T = 2\times 10^{-6}$, a typical dipolar encounter very weakly perturbs the thermal trajectory, but long-range interaction can still accumulate coherent spin-dependent phases over the evolution.
Direct internal-state-changing scattering is negligible.

\section*{Mean field approximation}

By decomposing the pairwise interaction
$
V_{ij}(t) = \bar{V} + \delta V_{ij}(t)
$
where
$
\bar{V} \equiv \langle V_{ij} \rangle\ \mathrm{and}\ \langle \delta V_{ij} \rangle = 0,
$
the average Hamiltonian becomes
$
\hat{H}_{\mathrm{coll}} = {\bar{V}}/{2} \sum_{i \neq j} \left( s_i^x s_j^x + s_i^y s_j^y \right).
$
Rewriting it in terms of the collective variables
$
\sum_{i \neq j} s_i^x s_j^x = S_x^2 - \sum_i \left( s_i^x \right)^2
$
such that
$
\sum_{i \neq j} \left( s_i^x s_j^x + s_i^y s_j^y \right) = S^2 - S_z^2 - {N}/{2},
$
the collective Hamiltonian is
\begin{equation}
    \hat{H}_{\mathrm{coll}} = \chi\, S^2 - \chi\, S_z^2 + \mathrm{const}
\end{equation}
with 
$
\chi = \bar{V}/2
$.
Note that $\chi S^2$ commutes with everything and, for a coherent spin state, it drives no dynamics.

For an initial state with population imbalance prepared at polar angle $\theta$,
$
\langle S_z \rangle = (N\cos\theta)/{2} .
$
Expand the OAT term using $S_z = \langle S_z \rangle + \delta S_z$,
\begin{equation}
    -\chi S_z^2 = -2\chi \langle S_z \rangle\, S_z + \chi \langle S_z \rangle^2 - \chi\, \delta S_z^2.
\end{equation}
In the mean-field approximation, we drop the second-order $\delta S_z^2$ and the constant $\langle S_z \rangle^2$ terms
\begin{equation}
    \hat{H}_{\mathrm{mf}} \approx  -2\chi \langle S_z \rangle\, S_z
\end{equation}
and hence
$
\omega_{\mathrm{mf}} = 2\chi \langle S_z \rangle = \chi N \cos\theta.
$

\end{document}


\title{Supplementary information: Parity-doublet spin interactions in ultracold polyatomic molecules}

\author{Junheng Tao}
\author{Paige Robichaud}
\author{Christian Hallas}
\author{Giseok Lee}
\author{John M. Doyle}
\affiliation{Department of Physics, Harvard University, Cambridge, MA 02138, USA}
\affiliation{Harvard-MIT Center for Ultracold Atoms, Cambridge, MA 02138, USA}

\date{\today}

\maketitle

\setcounter{secnumdepth}{3}
\renewcommand\theequation{S\arabic{equation}}
\renewcommand\thefigure{S\arabic{figure}}
\renewcommand\thetable{S\arabic{table}}
\renewcommand\thesection{S\arabic{section}}
\renewcommand\thesubsection{\thesection.\arabic{subsection}}

\section{Dipolar exchange interaction}

The two stretched parity-doublet states $|\!\uparrow\rangle\equiv|N=1,J=3/2,F=2,m_F=2,p=+\rangle$ and
$|\!\downarrow\rangle\equiv|N=1,J=3/2,F=2,m_F=2,p=-\rangle$ define a
pseudospin-$1/2$, with $s^z_i=(|\!\uparrow\rangle_i\langle\uparrow|
-|\!\downarrow\rangle_i\langle\downarrow|)/2$, raising and lowering
operators $s^+_i=|\!\uparrow\rangle_i\langle\downarrow|$,
$s^-_i=(s^+_i)^\dagger$, and transverse components
$s^x_i=(s^+_i+s^-_i)/2$, $s^y_i=-i(s^+_i-s^-_i)/2$. The two states share all
angular-momentum quantum numbers except parity and are split by the
parity-doublet gap $\Delta$ as shown in Fig.~1c.

Parity forbids a permanent dipole in either
state, $\langle\uparrow|\mathbf{d}|\uparrow\rangle=
\langle\downarrow|\mathbf{d}|\downarrow\rangle=0$, so at zero external
field the molecules carry no lab-frame dipole moment. The two states are
instead connected by a transition dipole
$d_{\uparrow\downarrow}=\langle\uparrow|\mathbf{d}|\downarrow\rangle
=0.73$~D. Because $|\!\uparrow\rangle$ and $|\!\downarrow\rangle$ share
the same projection $m_F=2$, the Wigner--Eckart theorem ($\Delta m_F=0$)
allows only the space-fixed $q=0$ (i.e.\ $z$) spherical component of
$\mathbf{d}$ between them. The dipole operator projected onto the
manifold is therefore off-diagonal in the parity basis and aligned with
the quantization axis,
\begin{equation}
\hat{\mathbf{d}}_i \;\longrightarrow\; d_{\uparrow\downarrow}\,
\hat{\mathbf{z}}\,\bigl(|\!\uparrow\rangle_i\langle\downarrow|
+|\!\downarrow\rangle_i\langle\uparrow|\bigr)
\;\propto\; 2d_{\uparrow\downarrow}\, s^x_i\,\hat{\mathbf{z}} .
\end{equation}

Two molecules separated by $\mathbf{r}$ interact through the electric dipole--dipole
Hamiltonian
\begin{equation}
H_{\rm dd}=\frac{1}{4\pi\epsilon_0 r^3}
\left[\hat{\mathbf{d}}_1\cdot\hat{\mathbf{d}}_2
-3(\hat{\mathbf{d}}_1\cdot\hat{\mathbf{r}})
(\hat{\mathbf{d}}_2\cdot\hat{\mathbf{r}})\right].
\end{equation}
With $\hat{\mathbf{z}}\cdot\hat{\mathbf{r}}=\cos\alpha$, where $\alpha$ is the angle between the intermolecular axis and the quantization field, projecting both dipoles into the parity-doublet manifold gives
\begin{equation}
\hat{H}_{\mathrm{dd}}^{(ij)} = \frac{d_{\downarrow}^{2}}{4\pi\epsilon_{0} r_{ij}^{3}}\left(1 - 3\cos^{2}\alpha_{ij}\right)\left(s_{i}^{+} + s_{i}^{-}\right)\left(s_{j}^{+} + s_{j}^{-}\right).
\end{equation}
After the rotating wave approximation
\begin{equation}
\hat{H}_{\mathrm{dd,RWA}}^{(ij)} = \frac{d_{\downarrow}^{2}}{4\pi\epsilon_{0} r_{ij}^{3}}\left(1 - 3\cos^{2}\alpha_{ij}\right)\left(s_{i}^{+}s_{j}^{-} + s_{i}^{-}s_{j}^{+}\right).
\end{equation}
Using $s_{i}^{+}s_{j}^{-} + s_{i}^{-}s_{j}^{+} = 2\left(s_{i}^{x}s_{j}^{x} + s_{i}^{y}s_{j}^{y}\right)$, the interaction becomes
$\hat{H}_{\mathrm{XY}} = \frac{1}{2}\sum_{i\neq j} V_{ij}\left(s_{i}^{x}s_{j}^{x} + s_{i}^{y}s_{j}^{y}\right)$ with
\begin{equation}
V_{ij} = \frac{2 d_{\downarrow}^{2}}{4\pi\epsilon_{0} r_{ij}^{3}}\left(1 - 3\cos^{2}\alpha_{ij}\right).
\end{equation}

\section{Collisional study}

\subsection{Collision Hamiltonian and basis}

The parity-doublet states are the opposite-parity stretched states
$
|\uparrow\rangle
=
|N=1,J=3/2,F=2,m_F=2,p=+\rangle,
$
and
$
|\downarrow\rangle
=
|N=1,J=3/2,F=2,m_F=2,p=-\rangle
$
with splitting
$
\nu_{\rm pd}=43\ {\rm MHz}.
$
At zero dc electric field, parity forbids a diagonal laboratory-frame dipole moment. Within the qubit manifold, the projected dipole operator is
\begin{equation}   
    P\widehat{\mathbf d}_iP
    =
    d_{\uparrow\downarrow}\widehat{\mathbf z}\,\sigma_i^x,
    \qquad
    d_{\uparrow\downarrow}=0.73\ {\rm D}.
\end{equation}
The projected two-molecule interaction is therefore
\begin{equation}   
\widehat V_{\rm dd}(R,\theta)
=
\frac{
d_{\uparrow\downarrow}^2
}{
4\pi\epsilon_0R^3
}
\left(
1-3\cos^2\theta
\right)
\sigma_1^x\sigma_2^x.
\end{equation}
In the Ramsey experiment, the initial state is symmetric upon particle exchange. 
Hence we choose the symmetric internal basis
$$
|q\rangle
\in
\left\{
|\Uparrow\rangle,
|T_0\rangle,
|\Downarrow\rangle
\right\},
$$
where
$$
|\Uparrow\rangle=|\uparrow\uparrow\rangle,
\qquad
|T_0\rangle=
\frac{
|\uparrow\downarrow\rangle+
|\downarrow\uparrow\rangle
}{\sqrt{2}},
\qquad
|\Downarrow\rangle=|\downarrow\downarrow\rangle.
$$
In this basis,
$$
\left.
\sigma_1^x\sigma_2^x
\right|_{\rm sym}
=
\begin{pmatrix}
0&0&1\\
0&1&0\\
1&0&0
\end{pmatrix}\equiv X.
$$
Taking the threshold of the mixed pair state as the zero of energy gives
$
{\epsilon_q}/{h}
=
\left(
+\nu_{\rm pd},
0,
-\nu_{\rm pd}
\right).
$

The channel basis is
\begin{equation}
|q;LM_L\rangle.
\end{equation}
As the internal pair state is symmetric, and CaOH is bosonic, only even partial waves occur. 
In atomic units, the channel Hamiltonian is
\begin{equation}
H_{qL,q'L'}^{(M_L)}(R)
=
\left[
\epsilon_q+
\frac{L(L+1)}{2\mu R^2}
\right]
\delta_{qq'}\delta_{LL'}
+
\frac{
d_{\uparrow\downarrow}^2
}{R^3}
X_{qq'}A_{LL'}^{(M_L)},
\end{equation}
with reduced mass
$
\mu=28.5\ {\rm u},
$
and angular matrix
$$
A_{LL'}^{(M)}
=
\int d\Omega\,
Y_{LM}^*(\Omega)
\left(
1-3\cos^2\theta
\right)
Y_{L'M}(\Omega)
$$
that couples $L$ and $L\pm2$.
We also define the dipole length and energy
\begin{equation}
a_d=\mu d_{\uparrow\downarrow}^2=4.29\times10^3a_0,
\qquad
E_d/k_B=0.165\ \mu{\rm K}.
\end{equation}
The large dipole length motivates both the high partial-wave cutoff and a large matching radius.

\subsection{Coupled-channel scattering calculation}

For total energy $E_{\mathrm{tot}}$, the matrix of radial solutions $F(R)$ obeys
\begin{equation}
\left[
-\frac{1}{2\mu}
\frac{d^2}{dR^2}\mathbf I
+
\mathbf H(R)
-
E_{\rm tot}\mathbf I
\right]
\mathbf F(R)
=0
\end{equation}
where $\mathbf H(R)$ is the channel Hamiltonian.
We propagate the logarithmic derivative
\begin{equation}
Y(R)=F'(R)F^{-1}(R)
\end{equation}
from the inner radius $R_0=30a_0$ to the asymptotic matching radius $R_m=10^4a_0$. 
These radii follow the numerical choices used in Ref.~\cite{vilas2026quantum}.

Similar to Ref.~\cite{vilas2026quantum}, we impose a universal loss short-range boundary condition at $R_0$~\cite{idziaszek2010universal}.
To do this, we diagonalize the channel Hamiltonian at
\begin{equation}
\mathbf H(R_0)\mathbf U
=
\mathbf U\,{\rm diag}(U_\alpha).
\end{equation}
The corresponding local radial wave numbers are
\begin{equation}
k_\alpha
=
\left[
2\mu
\left(
E_{\rm tot}-U_\alpha
\right)
\right]^{1/2}.
\end{equation}
For a locally open channel, a wave traveling from large to small has the form
\begin{equation}
F_\alpha(R)
\propto
\exp
\left[
-ik_\alpha(R-R_0)
\right].
\end{equation}
Its logarithmic derivative at the inner boundary is therefore
$
-ik_\alpha.
$
Transforming this condition back to the asymptotic channel basis gives
$
\mathbf Y(R_0)
=
\mathbf U\,
{\rm diag}(-ik_\alpha)
\mathbf U^\dagger.
$
For locally closed channels, the branch is chosen with
\begin{equation}
{\rm Re}\,k_\alpha\geq0,
\qquad
{\rm Im}\,k_\alpha\geq0.
\end{equation}

\subsection{Asymptotic matching}



















At $R_m$ the dipolar couplings are negligible compared with the channel kinetic energies.
We first partition $\mathbf{Y}$ into open and closed blocks and eliminate the asymptotically closed channels.
In the remaining open subspace $\mathbf{\tilde{Y}}_{oo}$, let $\mathbf H^-(R)$ and $\mathbf H^+(R)$ denote diagonal matrices of incoming and outgoing spherical Riccati-Hankel functions.
A scattering solution can be written as
\begin{equation}
\mathbf F_o(R)
=
\mathbf H^-(R)
-
\mathbf H^+(R)\mathbf S^{(q)}.
\end{equation}
Matching the logarithmic derivatives
\begin{equation}
\mathbf S^{(q)}
=
\left[
\mathbf H^{+\prime}
-
\widetilde{\mathbf Y}_{oo}\mathbf H^+
\right]^{-1}
\left[
\mathbf H^{-\prime}
-
\widetilde{\mathbf Y}_{oo}\mathbf H^-
\right].
\end{equation}
Primes denote radial derivatives.
Its flux-normalized matrix element
\begin{equation}
S_{q'L'M_L,qLM_L}^{(q)}
\end{equation} is the amplitude for an incoming channel $(q,L,M_L)$ to emerge in the open channel $(q',L',M_L)$.
Conservation of $M_L$ is written explicitly. 
The scattering matrix would be unitary if every outgoing channel were included and no short-range loss were present. 
Its missing norm is the probability of short-range capture.

\subsection{Partial-wave sums, cross sections, and rate coefficients}

We include every even incident and outgoing partial wave satisfying
\begin{equation}
|M_L|\leq L\leq L_{\max}.
\end{equation}
The calculations for $\pm M_L$ are degenerate so we propagate only nonnegative projections and account for the degeneracy using
\begin{equation}
g_M=
\begin{cases}
1,&M_L=0,\\
2,&M_L>0.
\end{cases}
\end{equation}
For an incident channel $i=(q,L,M_L)$, the probability missing from the open-channel scattering matrix is
\begin{equation}
P_i^{\rm cap}
=
1-
\sum_{f\in{\rm open}}
|S_{fi}|^2.
\end{equation}
The probability for inelastic scattering into a different retained internal pair
channel, excluding short-range loss, is
\begin{equation}
P_i^{\rm inel}
=
\sum_{\substack{
f\in{\rm open}\\
q_f\neq q
}}
|S_{fi}|^2.
\end{equation}

The initial product Ramsey state has pair-state populations
\begin{equation}
w_{\Uparrow}:w_{T_0}:w_{\Downarrow}
=
\frac14:\frac12:\frac14,
\end{equation}
the population-averaged cross sections are
\begin{equation}
\sigma_p(E_{\rm c})
=
\frac{2\pi}{k^2}
\sum_qw_q
\sum_{M_L\geq0}g_M
\sum_{\substack{
L\ {\rm even}\\
L\geq M_L
}}
P_{qLM_L}^{p}(E_{\rm c}),
\qquad
p\in
\left\{
{\rm cap},{\rm inel}
\right\}.
\end{equation}
Multiplying a cross section by the relative speed
\begin{equation}
v(E_{\rm c})
=
\frac{\hbar k}{\mu}
=
\sqrt{\frac{2E_{\rm c}}{\mu}}
\end{equation}
gives the corresponding two-body rate coefficient,
\begin{equation}
K_p(E_{\rm c})
=
v(E_{\rm c})\sigma_p(E_{\rm c}).
\end{equation}
These quantities use only probabilities and are sufficient for short-range capture and inelastic rate coefficients within the retained basis.
Ramsey coherence is phase sensitive: it requires the complex scattering amplitudes themselves, as described next.

\subsection{Scattering map}

A total elastic cross section measures how much outgoing flux differs from an
unscattered plane wave.
Ramsey contrast is reduced only if the outgoing motion, final internal state, or loss reservoir contains information about which component of the incoming spin superposition was present. 
We therefore construct an internal-state map
from the complex scattering amplitudes while retaining the labels of the
unobserved outgoing motional channels.

Consider an incoming internal pair state
\begin{equation}
|\chi\rangle
=
\sum_qc_q|q\rangle
\end{equation}
and an incident orbital channel
$
a=(L,M_L).
$
Each component enters with the same kinetic energy
$
E_{\rm c}.
$

The amplitude that contributes to an integrated scattering cross section is
proportional to the transition matrix
\begin{equation}
\mathbf T^{(q)}
=
\mathbf I-\mathbf S^{(q)}.
\end{equation}
For an incoming channel, the scattered state can be organized as
\begin{equation}
|\Psi_{\rm sc}^{(a)}\rangle
=
\sum_\lambda
\left[
A_{\lambda a}(E_{\rm c})
|\chi\rangle
\right]
\otimes
|\lambda\rangle.
\end{equation}
The label
\begin{equation}
\lambda=
\left(
E_{\rm out},L',M_L
\right)
\end{equation}
specifies an unobserved outgoing motional channel: its kinetic energy,
outgoing partial wave, and conserved orbital projection.

The cross-section-weighted operator acting on the internal pair space is
\begin{equation}
\left[
A_{\lambda a}(E_{\rm c})
\right]_{q'q}
=
\left(
\frac{2\pi g_M}{k^2}
\right)^{1/2}
\left[
\delta_{q'q}\delta_{L'L}
-
S_{q'L',qL}^{(q)}(E_{\rm c})
\right]
\chi_\lambda
\left(
E_{\rm c}+\epsilon_q-\epsilon_{q'}
\right).
\label{eq: A_matrix}
\end{equation}
The indicator enforcing equality of the outgoing kinetic energy and orbital channel is
\begin{equation}
\chi_\lambda(E)
=
\begin{cases}
1,&
E=E_{\rm out}
\ \text{and the outgoing orbital state is }(L',M_L),\\
0,&\text{otherwise}.
\end{cases}
\end{equation}
In the simulation, channels are placed in the same energy group when their outgoing kinetic energies agree.
Eq.~\eqref{eq: A_matrix} specifies which amplitudes must be added coherently.
For elastic scattering, $
E_{\rm out}=E_{\rm c}
$
for every component of the initial superposition. 
Those amplitudes therefore lead to the same outgoing motional state and must be combined coherently inside one operator $A_{\lambda a}$.
By contrast, amplitudes with different outgoing kinetic energies or orbital states multiply orthogonal motional states and are combined only after tracing over motion.

Tracing over the unresolved outgoing motion gives the scattered contribution to the internal pair density matrix,
\begin{equation}
\Phi_a^{\rm sc}[\rho]
=
\sum_\lambda
A_{\lambda a}\rho A_{\lambda a}^\dagger.
\label{eq: jump_map}
\end{equation}
Here,
$\rho$
is the incoming internal pair density matrix, and $\Phi_a^{\rm sc}$ denotes the scattered-state map for incident orbital channel $a$.

\subsection{Lindblad dissipator and short-range loss}

During a short interval $dt$, the probability for a molecule to encounter partners in a cross section $\sigma$ is
\begin{equation}
n\,v(E_{\rm c})\,\sigma\,dt,
\end{equation}
where $n$ is the local density. 
Combining the jump map in Eq.~\eqref{eq: jump_map} with the corresponding no-jump amplitude gives
\begin{equation}
\rho(t+dt)-\rho(t)
=
n\,dt\,
\mathcal D_{E_{\rm c}}[\rho]
+
O(dt^2),
\end{equation}
where
\begin{equation}
\mathcal D_{E_{\rm c}}[\rho]
=
v(E_{\rm c})
\sum_{a,\lambda}
\left[
A_{\lambda a}\rho A_{\lambda a}^\dagger
-
\frac12
\left\{
A_{\lambda a}^\dagger A_{\lambda a},
\rho
\right\}
\right].
\label{eq: dissipator}
\end{equation}

A complete scattering master equation also contains coherent
forward scattering. We denote by
\begin{equation}
W_{\rm fs}(E_{\rm c})
\end{equation}
the Hermitian pair-spin operator obtained from the real part of the zero-angle scattering amplitude.
Its contribution is
\begin{equation}
\left.
\frac{d\rho}{dt}
\right|_{\rm fs}
=
-\frac{i}{\hbar}
\left[
nW_{\rm fs}(E_{\rm c}),
\rho
\right].
\end{equation}
This term produces a coherent density-dependent phase or frequency shift. 
We do not include it in the reported Ramsey contrast-decay rate coefficient,
because coherent long-range spin evolution is already propagated by DTWA. The
dissipator in Eq.~\eqref{eq: dissipator} is also not added to the DTWA rate: in the
weak-deflection regime, averaging spin-dependent scattering phases over
unresolved outgoing channels and averaging the same interaction over sampled
classical trajectories are overlapping descriptions.

The open-channel matrix is subunitary because the inner boundary absorbs flux.
We define a matrix of short-range loss cross sections $\boldsymbol{\Sigma}_{\rm loss}$, whose diagonal element is the total capture cross section for an
incoming internal pair state
\begin{equation}
\left[
\boldsymbol{\Sigma}_{\rm loss}
\right]_{qq}
=
\sum_a
\frac{2\pi g_M}{k^2}
\left[
1-
\sum_{f\in{\rm open}}
\left|
S_{f,qa}^{(q)}
\right|^2
\right].
\end{equation}

The capture contribution to the unnormalized surviving pair state is
\begin{equation}
\mathcal L_{E_{\rm c}}^{\rm cap}[\rho]
=
-\frac{
v(E_{\rm c})
}{2}
\left\{
\boldsymbol{\Sigma}_{\rm loss},
\rho
\right\}.
\end{equation}
The energy-resolved evolution used to calculate the reported initial decay is
therefore

\begin{equation}
\left.
\frac{d\rho}{dt}
\right|_{E_{\rm c}}
=
n
\left[
\mathcal D_{E_{\rm c}}[\rho]
+
\mathcal L_{E_{\rm c}}^{\rm cap}[\rho]
\right].
\end{equation}

\subsection{Ramsey contrast decay}

The initial pair state is
\begin{equation}
|\psi_0\rangle
=
|+x,+x\rangle
=
\frac12
\left(
|\Uparrow\rangle
+
\sqrt2|T_0\rangle
+
|\Downarrow\rangle
\right),
\qquad
\rho_0=|\psi_0\rangle\langle\psi_0|.
\end{equation}
Write the surviving pair density matrix in the ordered basis
$
\left(
\Uparrow,T_0,\Downarrow
\right),
$
and tracing over the second molecule gives
\begin{equation}
\rho^{(1)}
=
{\rm Tr}_2\rho
=
\begin{pmatrix}
\rho_{\Uparrow\Uparrow}
+
\rho_{T_0T_0}/2
&
\left(
\rho_{\Uparrow T_0}
+
\rho_{T_0\Downarrow}
\right)/\sqrt2
\\
\left(
\rho_{T_0\Uparrow}
+
\rho_{\Downarrow T_0}
\right)/\sqrt2
&
\rho_{\Downarrow\Downarrow}
+
\rho_{T_0T_0}/2
\end{pmatrix}.
\end{equation}
Its trace
\begin{equation}
N_s={\rm Tr}\rho^{(1)}={\rm Tr}\rho
\end{equation}
is the surviving norm. 
The normalized Ramsey contrast along $x$ is
\begin{equation}
C_x
=
\frac{
{\rm Tr}
\left[
\sigma_x\rho^{(1)}
\right]
}{
N_s
}
=
\frac{
2\,{\rm Re}
\left[
\rho_{\uparrow\downarrow}^{(1)}
\right]
}{
{\rm Tr}\rho^{(1)}
}.
\end{equation}
Differentiating the observable
\begin{equation}
\frac{dC_x}{dt}
=
\frac{
2\,{\rm Re}
\left[
\dot\rho_{\uparrow\downarrow}^{(1)}
\right]
}{
N_s
}
-
C_x
\frac{\dot N_s}{N_s}.
\end{equation}
The second term is essential when loss is present. 
Take $t=0$
\begin{equation}
\left.
\frac{dC_x}{dt}
\right|_{t=0}
=
2\,{\rm Re}
\left[
\dot\rho_{\uparrow\downarrow}^{(1)}
\right]_{t=0}
-
\dot N_s(0).
\end{equation}

The familiar phase-shift result is recovered as a special limit. 
Suppose the
scattering matrix is diagonal in all pair-spin channels, there is no
loss, and only one outgoing motional state is relevant
\begin{equation}
S_q=e^{2i\delta_q}.
\end{equation}
where $\delta_q$ is the real elastic scattering phase shift in pair channel $q$.
Directly applying these phases and tracing over one molecule gives
\begin{equation}
C_x'
=
\cos
\left(
\delta_{\Downarrow}-\delta_{\Uparrow}
\right)
\cos
\left(
\delta_{\Downarrow}
+
\delta_{\Uparrow}
-
2\delta_{T_0}
\right).
\end{equation}
When the two aligned channels have the same phase,
$
\delta_{\Uparrow}
=
\delta_{\Downarrow}
\equiv
\delta_\parallel,
$
the contrast change per collision becomes
\begin{equation}
\Delta C
=
1-C_x'
=
2\sin^2
\left(
\delta_\parallel-\delta_{T_0}
\right).
\end{equation}
This is the single-phase-difference expression used in simplified collision
models~\cite{carroll2025observation, wang2025theory}. 
The full scattering map is required here because the CaOH scattering
matrix can couple orbital and internal channels, so a single diagonal phase shift is not sufficient.

Finally, we define the energy-resolved initial Ramsey contrast-decay rate
coefficient
\begin{equation}
K_R(E_{\rm c})
=
-\frac{1}{nC_x(0)}
\left.
\frac{dC_x}{dt}
\right|_{t=0,E_{\rm c}}.
\end{equation}

\subsection{Thermal and trap averaging}

For a nondegenerate three-dimensional gas, the relative collision-energy
distribution is
\begin{equation}
p_T(E)
=
\frac{2}{\sqrt{\pi}}
\frac{\sqrt{E}}{(k_BT)^{3/2}}
\exp
\left(
-\frac{E}{k_BT}
\right).
\end{equation}
The thermal rate coefficients are
\begin{equation}
\left\langle K_p\right\rangle_T
=
\int_0^\infty
dE\,
p_T(E)K_p(E),
\qquad
p\in
\left\{
R,{\rm cap},{\rm inel}
\right\}.
\end{equation}
Energy-correlated quantities are averaged before any factorization. 
For example,
\begin{equation}
\left\langle K\Delta C\right\rangle_T
=
\int_0^\infty
dE\,
p_T(E)
K(E)\Delta C(E),
\end{equation}

For a Gaussian density profile, the initial density experienced by a
randomly selected molecule is
\begin{equation}
\overline n
=
\frac{
\int d^3\mathbf r\,n^2(\mathbf r)
}{
\int d^3\mathbf r\,n(\mathbf r)
}
=
\frac{n_0}{2^{3/2}},
\end{equation}

where $n_0$ is the peak density. 
The initial cloud-averaged rate is therefore 
\begin{equation}
\Gamma_p(0)
=
\overline n
\left\langle K_p\right\rangle_T.
\end{equation}

\subsection{Simulations and results}

We evaluated the scattering calculation described above without fitting any collision parameter to the Ramsey data. 
For each collision energy \(E_c\), we propagated the three distinct incoming internal thresholds \(q\in\{\Uparrow,T_0,\Downarrow\}\) in every \(M_L\) block. 
Within each block, all even incident and outgoing partial waves satisfying
$
|M_L|\leq L\leq L_{\max}
$
were included.
A single propagation at fixed \(q\), \(M_L\), and \(E_c\) supplies all incident-\(L\) columns of that block.

The fundamental radial solution and its derivative were integrated on 120
equal radial segments between \(R_0=30a_0\) and \(R_m=10^4a_0\). 
Within each segment we used an eighth-order adaptive Dormand-Prince integrator with relative and absolute tolerances \(10^{-8}\) and \(10^{-10}\), respectively. 

Closed-channel solutions contain exponentially growing components that can make the fundamental solution matrix \(\mathbf F\) ill-conditioned. After each radial segment, we factorize \(\mathbf F=\mathbf Q\mathbf R\) and replace \((\mathbf F,\mathbf F')\) by \((\mathbf Q,\mathbf F'\mathbf R^{-1})\). 
This leaves the log derivative invariant
$
(\mathbf F'\mathbf R^{-1})(\mathbf F\mathbf R^{-1})^{-1}
=\mathbf F'\mathbf F^{-1},
$
while restoring orthonormal columns and preventing exponentially growing closed-channel components from erasing the numerically smaller independent solutions.
The calculation used the parameters in Table S1.

\begin{table}[htb]
\label{tab:scattering-params}
\begin{ruledtabular}
\begin{tabular}{lr}
\textit{Parameters} & \textit{Values} \\ \hline
reduced mass ($\mu$)                      & $28.5\,\mathrm{u}$ \\
parity-doublet splitting ($\nu_{\rm pd}$) & $43\,\mathrm{MHz}$ \\
transition dipole         & $0.73\,\mathrm{D}$ \\
inner boundary ($R_0$)                    & $30\,a_0$ \\
matching radius ($R_m$)                   & $10^{4}\,a_0$ \\
radial QR segments                        & 120 \\
thermal partial waves ($L_{\max}$)    & 20 \\
collision-energy grid & 12 logarithmic points, $2$--$600\,\mu\mathrm{K}$ \\
temperature                               & $60\,\mu\mathrm{K}$ \\
short-range boundary                      & universal absorption loss \\
\end{tabular}
\end{ruledtabular}
\caption{Scattering calculation parameters.}
\end{table}

For each energy, the complex amplitudes were assembled into the operators
\(A_{\lambda a}\). 
We grouped amplitudes coherently only when they had the same outgoing kinetic energy, \(L'\), and \(M_L\).
The resulting dissipator was applied to the initial pair state, and the
derivative of the normalized one-body contrast was evaluated.

With these we obtain
\[
\langle K_R\rangle_T
=1.67\times10^{-9}\ {\rm cm^3\,s^{-1}}
\]
and
\[
\langle K_{\rm cap}\rangle_T
=4.83\times10^{-10}\ {\rm cm^3\,s^{-1}}.
\]
The direct internal-state-changing coefficient among the parity-doublet channels is approximately
\[
\langle K_{\rm inel}\rangle_T
\simeq9.5\times10^{-14}\ {\rm cm^3\,s^{-1}}.
\]
The Ramsey coefficient arises almost entirely from spin-dependent elastic scattering where different spin components produce distinguishable outgoing motional states.
The direct inelastic collision contribution is negligible.

\begin{figure}[tb!]
\centering
\includegraphics{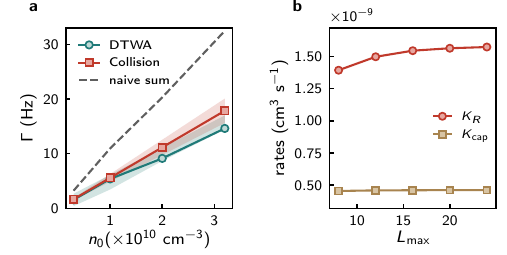}
\caption{Two-body scattering prediction for Ramsey contrast decay and numerical convergence.
{\bf a} Interaction-induced Ramsey decay rate \(\Gamma\) versus peak molecular density \(n_0\). Green circles show the many-body DTWA prediction, and red squares show the initial rate obtained from the two-body collision after thermal averaging.
Shaded bands denote the propagated uncertainties of the respective calculations. 
The grey dashed line is their naive sum, which substantially overpredicts the decay; The two approaches contain overlapping descriptions of long-range dipolar dephasing.
{\bf b} Partial-wave convergence at \(E_c/k_B=60\,\mu{\rm K}\) of the Ramsey-decay coefficient \(K_R\) and number loss capture coefficient \(K_{\rm cap}\). The Ramsey coefficient approaches convergence with increasing \(L_{\max}\), while \(K_{\rm cap}\) is stable.}
\label{fig:SM_collision}
\end{figure}

The collision model and DTWA rates agree closely within uncertainty, as shown in Fig.~\ref{fig:SM_collision}.
The predicted molecule number loss rate is also close to the experimental observations and that in Ref.~\cite{vilas2026quantum}. 
This is an important consistency check, but it does not identify an additional decay
channel that should be added to DTWA. Both calculations use the same
\(1/r^3\) parity-doublet interaction and the same thermal relative motion.
DTWA samples the coherent spin phase along classical trajectories and averages
over those trajectories. The scattering calculation encodes the same
two-body evolution in asymptotic motional channels and traces over the outgoing
motion. In the present weak-deflection semiclassical regime, these are
overlapping descriptions of long-range dipolar dephasing.

For illustration, adding the two interaction-induced rates would substantially exceed either independent calculation as well as experimental data. 
We therefore use DTWA as
the many-body prediction in the main text and the collision model as an
independent two-body cross-check. 
A combined calculation would require a
matched kinetic treatment that separates the coherent self-energy from the
dissipative collision integral, or a joint propagation of spin and motion.

\bibliography{main}